\documentclass[a4paper,onecolumn,11pt,accepted=2023-12-11]{quantumarticle}
\pdfoutput=1
\usepackage[utf8]{inputenc}
\usepackage[english]{babel}
\usepackage[T1]{fontenc}
\usepackage{amsmath}
\usepackage{hyperref}

\usepackage{amssymb}
\usepackage{bbm}
\usepackage{mathtools}

\usepackage{tikz}
\usepackage{lipsum}

\newcommand{\tr}[1]{\operatorname{Tr}\left( #1 \right)}
\newcommand{\norm}[1]{\lVert#1\rVert}
\newcommand{\diag}[1]{\operatorname{diag}\left(#1\right)}
\newcommand{\Vecc}[1]{\operatorname{Vec}\left(#1\right)}

\newcommand{\qdim}{d}

\DeclareMathOperator*{\argmin}{arg\,min}
\newcommand{\bra}[1]{\langle #1 |}
\newcommand{\ket}[1]{| #1 \rangle}

\newcommand{\dm}[1]{ | #1 \rangle \langle #1|}
\newcommand{\kb}[2]{ | #1 \rangle \langle #2|}

\newcommand{\I}{\mathbb{I}} 
\newcommand{\cond}[1]{\mathbbm{1}\lbrace #1 \rbrace} 
\newcommand{\E}{\mathbb{E}} 

\newcommand{\X}{\mathcal{X}}

\newcommand{\C}{\mathbb{C}} 
\newcommand{\Z}{\mathbb{Z}} 
\newcommand{\R}{\mathbb{R}} 
\newcommand{\U}[1]{\operatorname{U}\left( #1 \right)} 

\newcommand{\y}{\mathbf{y}} 
\newcommand{\bx}{\mathbf{x}} 

\newcommand{\balpha}{\boldsymbol{\alpha}}
\newcommand{\bnu}{\boldsymbol{\nu}}

\usepackage{makecell}
\usepackage{amsthm}
\theoremstyle{definition}

\newtheorem*{theorem*}{Theorem}

\newtheorem*{lemma*}{Lemma}

\newtheorem*{corollary*}{Corollary}

\newcommand{\fweight}[1]{feature weight#1} 
\newcommand{\fcoef}[1]{Fourier coefficient#1} 
\newcommand{\tcoef}[1]{trainable weight#1}

\begin{document}

\title{Generalization despite overfitting in quantum machine learning models}

\author{Evan Peters}
\email{e6peters@uwaterloo.ca}

\affiliation{Department of Physics, University of Waterloo, Waterloo, ON, N2L 3G1, Canada}
\affiliation{Institute for Quantum Computing, Waterloo, ON, N2L 3G1, Canada}
\affiliation{Perimeter Institute for Theoretical Physics, Waterloo, Ontario, N2L 2Y5, Canada}

\author{Maria Schuld}
\affiliation{Xanadu, Toronto, ON, M5G 2C8, Canada}
\maketitle

\begin{abstract}
The widespread success of deep neural networks has revealed a surprise in classical machine learning: very complex models often generalize well while simultaneously overfitting training data. This phenomenon of benign overfitting has been  studied for a variety of classical models with the goal of better understanding the mechanisms behind deep learning. Characterizing the phenomenon in the context of quantum machine learning might similarly improve our understanding of the relationship between overfitting, overparameterization, and generalization. In this work, we provide a characterization of benign overfitting in quantum models. To do this, we derive the behavior of a classical interpolating Fourier features models for regression on noisy signals, and show how a class of quantum models exhibits analogous features, thereby linking the structure of quantum circuits (such as data-encoding and state preparation operations) to overparameterization and overfitting in quantum models. We intuitively explain these features according to the ability of the quantum model to interpolate noisy data with locally ``spiky'' behavior and provide a concrete demonstration example of benign overfitting.
\end{abstract}

 \section{Introduction}
 
 A long-standing paradigm in machine learning is the trade-off between the complexity of a model family and the model's ability to generalize: more expressive model classes contain better candidates to fit complex trends in data, but are also prone to overfitting noise \cite{nielsen2015neural,geman1992}. \textit{Interpolation}, defined for our purposes as choosing a model with zero training error, was hence long considered bad practice \cite{hastie2009elements}. The success of deep learning -- machine learning in a specific regime of extremely complex model families with vast amounts of tunable parameters -- seems to contradict this notion; here, consistent evidence shows that among some interpolating models, more complexity tends  \textit{not to harm} the generalisation performance\footnote{The fact that interpolation does not always harm generalization performance is in fact well-known for models like boosting and nearest neighbours, which are both interpolating models.}, a phenomenon described as ``benign overfitting'' \cite{bartlett2021deep}. 

\begin{figure}
    \centering
    \includegraphics[width=\textwidth]{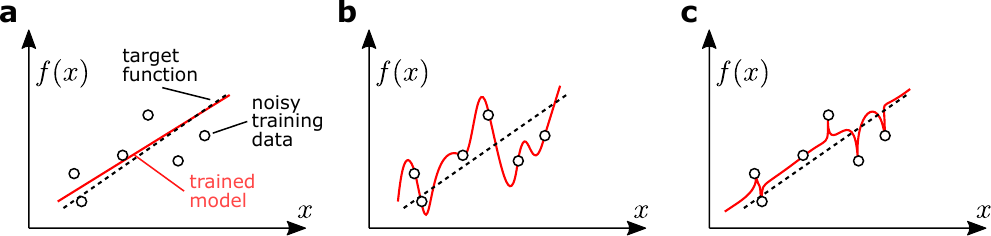}
    \caption{Intuition behind the phenomenon of benign overfitting with spiking models. \textbf{a} Training a model typically involves a trade-off between fitting noisy training data well and recovering an underlying target function \textbf{b} Traditional learning theory associates interpolating models (reaching zero training error by fitting every data point) with low generalization capability (or high test error) by failing to recover a simple target function. \textbf{c} ``spiking models'' that change quickly in the vicinity of training data but otherwise exhibit simple behavior can explain how both kinds of errors may be kept low.}
    \label{fig:benign_overfit}
\end{figure}

In recent years, a surge of theoretical studies have reproduced benign overfitting in simplified settings with the hope of isolating the essential ingredients of the phenomenon \cite{bartlett2021deep,belkin2021fit}. For example, Ref.~\cite{bartlett2020benign} showed how interpolating linear models in a high complexity regime (more dimensions than datapoints) could generalize just as well as their lower-complexity counterparts on new data, and analyzed the properties of the data that lead to the ``absorption'' of noise by the interpolating model without harming the model's predictions. Ref.~\cite{belkin_reconcile} showed that there are model classes of simple functions that change quickly in the vicinity of the noisy training data, but recover a smooth trend elsewhere in data space (see Figure~\ref{fig:benign_overfit}). Such functions have also been used to train nearest neighbor models that perfectly overfit training data while generalizing well, thereby directly linking ``spiking models'' to benign overfitting~\cite{belkin2019does}. Recent works try to recover the basic mechanism of such spiking models using the language of Fourier analysis~\cite{muthukumar2020harmless,muthukumar2021classification,dar2021}.  

In parallel to these exciting developments in the theory of deep learning, quantum computing researchers have proposed families of parametrised quantum algorithms as model classes for machine learning (e.g. Ref.~\cite{Benedetti_2019}). These quantum models can be optimised similarly to neural networks \cite{Mitarai_2018,Schuld_2019} and have interesting parallels to kernel methods \cite{schuld2019b,havlicek2019} and generative models \cite{Lloyd_2018,Dallaire_Demers_2018}. Although researchers have taken some first steps to study the expressivity \cite{Abbas_2021,Wright_20,Sim_2019,hubregsten2021}, trainability \cite{mcclean2018barren,cerezo2021cost} and generalisation  \cite{caro2021,huang2021power,caro2021generalization,PRXQuantum.2.040321} of quantum models, we still know relatively little about their behaviour. In particular, the interplay of overparametrisation, interpolation, and generalisation that seems so important for deep learning is yet largely unexplored. 

In this paper we develop a simplified framework in which questions of overfitting in quantum machine learning can be investigated. Essentially, we exploit the observation that quantum models can often be described in terms of Fourier series where well-defined components of the quantum circuit influence the selection of Fourier modes and their respective Fourier coefficients \cite{gil2020input, schuld2021, wierichs2022general}. We link this description to the analysis of spiking models and benign overfitting by building on prior works analyzing these phenomena using Fourier methods. In this approach, the complexity of a model is related to the number of Fourier modes that its Fourier series representation consists of, and overparametrised model classes have more modes than needed to interpolate the training data (i.e., to have zero training error). After deriving the generalization error for such model classes these ``superfluous'' modes lead to spiking models, which have large oscillations around the training data while keeping a smooth trend everywhere else. However, large numbers of modes can also harm the recovery of an underlying signal, and we therefore balance this trade-off to produce an explicit example of benign overfitting in a quantum machine learning model.

The mathematical link described above allows us to probe the impact of important design choices for a simplified class of quantum models on this trade-off. For example, we find why a measure of redundancy in the spectrum of the Hamiltonian that defines standard data encoding strategies strongly influences this balance; in fact to an extent that is difficult to counterbalance by other design choices of the circuit.   

The remainder of the paper proceeds as follows. We will first review the classical Fourier framework for the study of interpolating models and develop explicit formulae for the error in these models to produce a basic example of benign overfitting (Sec.~\ref{sec:class}). We will then construct a quantum model with analogous components to the classical model, and demonstrate how each of these components is related to the structure of the corresponding quantum circuit and measurement (Sec.~\ref{sec:qmodel}). We then analyze specific cases that give rise to ``spikiness'' and benign overfitting in these quantum models (Sec.~\ref{sec:qmodel2}).

\section{Interpolating models in the Fourier framework}\label{sec:class}

In this section we will provide the essential tools to probe the phenomenon of overparametrized models that exhibit the spiking behaviour from Figure~\ref{fig:benign_overfit} using the language of Fourier series. We will review and formalize the problem setting and several examples from Refs.~\cite{muthukumar2020harmless,muthukumar2021classification,dar2021}, before extending their framework by incorporating standard results from linear regression to derive closed-form error behavior and examples of benign overfitting.

\subsection{Setting up the learning problem}

We are interested in functions $g$ defined on a finite interval that may be written in terms of a linear combination of Fourier basis functions or \textit{modes} $e^{i 2\pi k x}$ (each describing a complex sinusoid with integer-valued frequency $k$) weighted by their corresponding \fcoef{s} $\hat{g}_k$:
\begin{equation}
    g(x) = \sum_{k=-\infty}^\infty   \hat{g}_k e^{i 2\pi k x}.
\end{equation}

We restrict our attention to well-behaved functions that are sufficiently smooth and continuous to be expressed in this form. We will now set up a simple learning problem whose basic components -- the model and target function to be learned -- can be expressed as Fourier series with only few non-zero coefficients, and define concepts such as overparametrization and interpolation. 

Consider a machine learning problem in which data is generated by a target function of the form
\begin{equation}\label{eq:bandlimited}
    g(x) = \sum_{k\in \Omega_{n_0}} \hat{g}_k e^{i 2\pi k x}
\end{equation}
which only contains frequencies in the discrete, integer-valued spectrum
\begin{equation}\label{eq:spec_n}
\Omega_{n_0} = \Bigl\{ -\frac{n_0-1}{2}, -\frac{n_0-1}{2} + 1, \dots, 0, \dots, \frac{n_0-1}{2} \Bigr\}
\end{equation}
for some odd integer $n_0$. We call functions of this form \textit{band-limited} with bandwidth $n_0/2$ (that is, $|k| < n_0/2$ for all frequencies $k \in \Omega_{n_0}$). The bandwidth limits the complexity of a function, and will therefore be important for exploring scenarios where a complex model is used to overfit a less complex target function. We are provided with $n$ noisy samples $y_j = g(x_j) + \epsilon$ of the target function $g$ evaluated at points $x_j=j/n$ spaced uniformly on interval $[0, 1]$ (we will assume input data has been rescaled to $[0, 1]$ without loss of generality), where $n > n_0$ and we require $n$ to be odd for simplicity.  

The model class we consider likewise contains band-limited Fourier series, and since we are interested in interpolating models, we always assume that they have enough modes to interpolate the noisy training data, namely:
\begin{align}\label{eq:model}
    f(x) &= \sum_{k\in \Omega_d} \alpha_k \sqrt{\nu_k} e^{i 2\pi k x}, \\ 
    \text{such that }\qquad f(x_j) &= y_j \text{ for all } j\in[n]. \nonumber
\end{align}
Similarly to Eq.~\eqref{eq:spec_n} we define the spectrum
\begin{equation}
    \Omega_d = \Bigl\{ -\frac{d-1}{2}, -\frac{d-1}{2} + 1, \dots, 0, \dots, \frac{d-1}{2} \Bigr\}.
\end{equation}

Following Ref. \cite{muthukumar2020harmless}, this model class has two components: The set of weighted Fourier basis functions $\sqrt{\nu_k} e^{i2\pi k x}$ describe a feature map applied to $x$ for some set of \fweight{s} $\nu_k \in \R^+$, while the \tcoef{s} $\alpha_k\in \C$ are optimized to ensure that $f$ interpolates the data. The theory of trigonometric polynomial interpolation \cite{atkinson2008introduction} ensures that $f$ can always interpolate the training data for some choice of \tcoef{s} $\alpha_k$ under these conditions. In the following, we will therefore usually consider the $\alpha_k$ as being determined by the data and interpolation condition, while the $\nu_k$ serve as our ``turning knobs'' to create different settings in which to study spiking properties and benign overfitting. We call the model class described in Eq.~\eqref{eq:model} \textit{overparameterized} when the degree of the Fourier series is much larger than the degree of the target function, $d \gg n_0$, in which case the model has many more frequencies available to perform fitting than there are present in the underlying signal to fit. 

Note that one can rewrite the Fourier series of Eq.~\eqref{eq:model} in a linear form
\begin{equation}\label{eq:f_dot}
    f(x) = \langle \boldsymbol{\alpha}, \phi(x)\rangle
\end{equation}
where 
\begin{align}
    \boldsymbol{\alpha} &= (\alpha_{-(d-1)/2}, \dots, \alpha_k, \dots, \alpha_{(d-1)/2}) \in \C^d,
    \\ \phi(x) &= \left(\sqrt{\nu_{-(d-1)/2}} e^{i\pi (d-1)x} \dots, \sqrt{\nu_k}e^{i 2\pi kx}, \dots,  \sqrt{\nu_{(d-1)/2}} e^{i\pi (d-1)x} \right) \in \C^d. \label{eq:weighted_ff}
\end{align}
From this perspective, optimizing $f$ amounts to learning \tcoef{s} $\boldsymbol{\alpha}$ by performing regression on observations $\phi(x)$ sampled from a \textit{random Fourier features} model \cite{rahimi2007random} for which $\nu_k$ and $\phi(x)_k$ are precisely the eigenvalues and eigenfunctions of a shift-invariant kernel \cite{rudin}.

To complete the problem setup, we have to impose one more constraint. Consider that $\exp(i 2\pi k x_j)$ with frequency $k < n/2$ and uniformly spaced points $x_j$ is equal to $\exp(i 2\pi k' x_j)$ for any choice of \textit{alias frequency} $k' = k \mod n$. The presence of these aliases means that the model class described in Eq.~\eqref{eq:f_dot} contains many interpolating solutions in the overparameterized regime. Motivated by prior work exploring benign overfitting  for linear features \cite{bartlett2020benign}, Fourier features \cite{muthukumar2020harmless,mei2022}, and other nonlinear features \cite{hastie2022surprises,liang2020multiple}, we will study the minimum-$\ell_2$ norm interpolating solution,
\begin{equation}\label{eq:classical_problem}
    \boldsymbol{\alpha}^{opt} = \argmin_{\boldsymbol{\alpha}: f(x_j) = y_j \, \forall \, j} \norm{\boldsymbol{\alpha}}_2.
\end{equation}

Minimizing the $\ell_2$ norm is a typical choice for imposing a penalty on complex functions (regularization) in the underparameterized regime, though we will see that this intuition does not carry over to the overparameterized regime. The remainder of this section will explore how this learning problem results in a trade-off in interpolating Fourier models: Overparameterization introduces alias frequencies that increase the error in fitting simple target functions but can also reduce error by absorbing noise into high-frequency modes with spiking behavior.

\subsection{Two extreme cases to understand generalization}\label{sec:cases}

To better understand the trade-off that overparametrization -- or in our case, a much larger number of Fourier modes than needed to interpolate the data -- introduces between fitting noise and generalization error, we revisit two extreme cases explored in Ref.~\cite{muthukumar2020harmless}, involving a pure-noise signal and a noiseless signal. 

\subsubsection*{Case 1: Noise only}

\begin{figure}
    \centering
    \includegraphics[width=\textwidth]{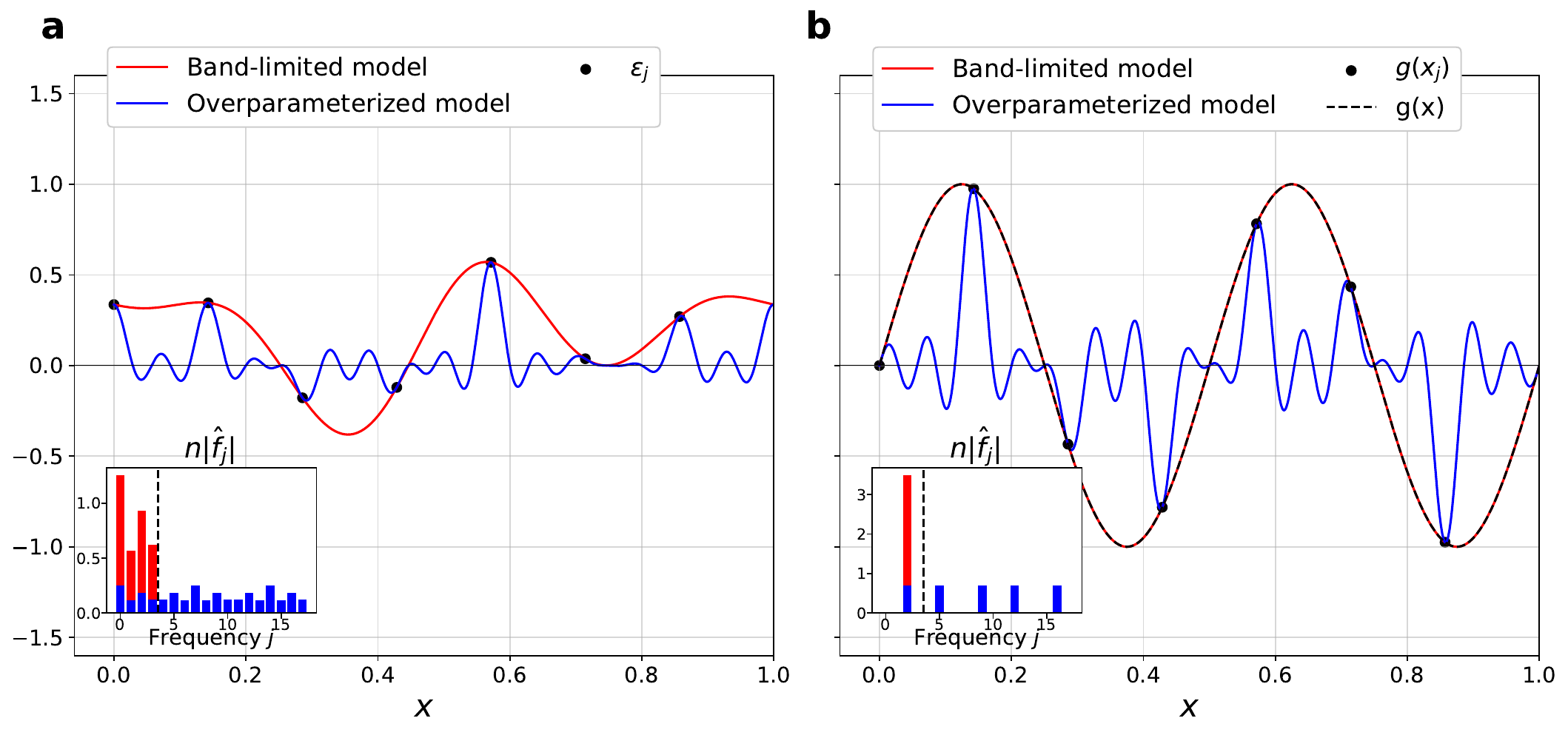}
    \caption{Comparison of overparametrized and simple models that interpolate data from different target functions. \textbf{a} \textit{Noise only:} The overparameterized model ($n=7,\, d=35$, blue) is more effective at recovering the target function $g(x)=0$ when provided with noisy data, while a band-limited model (red) cannot do so effectively. Inset: The distribution of \tcoef{s} for the minimum $\norm{\boldsymbol{\alpha}}_2$ model are distributed among many aliases suppressing the error of the optimized model (proportional to $\sum_j |\hat{f}_j|^2$) in this case. \textbf{b} \textit{Signal only:} The opposite occurs in the case of noiseless input, for which a band-limited model ($n_0 = 5$) perfectly recovers $g(x) = \sin\left(2 \pi (2x)\right)$ while the overparameterized model fails to capture the behavior of the input signal. Inset: Minimizing the $\ell_2$ norm of $\boldsymbol{\alpha}$ distributes \fcoef{s} $\hat{g}_2$ among aliases $\hat{f}_{2+\ell n}$ (and $\hat{g}_{-2}$ among aliases $\hat{f}_{-2 + \ell n}$); this bleeding of the signal $g$ into higher frequencies results in higher error in the overparameterized model  \cite{muthukumar2020harmless,dar2021}.}
    \label{fig:noise_noiseless_demo}
\end{figure}

The first case demonstrates how alias modes can help to fit noise without disturbing the (here trivial) signal. We set $g=0$ and consider $n$ observations $y_j = \epsilon_j$ of zero-mean noise with known variance $\E[\epsilon^2] = \sigma^2$. After making $n$ uniformly spaced observations, we compute the discrete Fourier transform of the observations as the sequence of values $\hat{\epsilon}_j$ satisfying
\begin{equation}
    \hat{\epsilon}_j = \frac{1}{n}\sum_{k \in \Omega_n} \epsilon_k e^{-i 2\pi jk/n},
\end{equation}
which characterizes the frequency content of the noisy signal that can be captured and learned from using only $n$ evenly spaced samples. Suppose that the degree of the model (controlling the model complexity) is given by $d = n(m+1)$ for some even integer $m$ and that $\nu_k = 1$ for every mode, so that there are exactly $m$ equally-weighted aliases for each frequency in the spectrum of the Fourier series for $g$. Then the optimal (i.e., the minimum $\ell_2$-norm, interpolating) trainable weight vector $\boldsymbol{\alpha}^{opt}$ has entries
\begin{equation}\label{eq:noiseonly_sol}
    \alpha^{opt}_{k + n\ell} = \frac{\hat{\epsilon}_k}{n(m+1)}
\end{equation}
for $\ell = -m/2, \dots, m/2$, with all other entries being zero (see Appendix~\ref{app:special_cases}). From Eq.~\eqref{eq:noiseonly_sol}, the minimum-$\norm{\boldsymbol{\alpha}}_2$ solution distributes noise Fourier coefficients $\hat{\epsilon}_k$ evenly into many alias frequencies $k+n\ell$, while enforcing that the sum of \tcoef{s} $\alpha_{k+n\ell}$ for all of these aliases is $\hat{\epsilon}_k$ to guarantee interpolation. As shown in Figure~\ref{fig:noise_noiseless_demo}, the higher-frequency aliases suppress the optimal model $f^{opt}(x) = \langle \boldsymbol{\alpha}^{opt}, \phi(x)\rangle$ to near-zero at every point away from the interpolated points, resulting in a test error of $\mathcal{O}(\sigma^2 / m)$ that decreases monotonically with the complexity of the model. As $m$ increases, the optimal model $f^{opt}$ remains close to the true signal $y=0$ while becoming ``spiky'' near the noisy samples. By conforming to the true signal everywhere except in the vicinity of noise, this behavior embodies the mechanism of how overparameterized models can absorb noise into high frequency modes. In this case the generalization error, measuring how close the model is to the target function on average, \textit{decreases} with increasing complexity of the model class.

\subsubsection*{Case 2: Signal only}

While the above case shows how overparametrization can help to absorb noise to reduce error without harming the signal, the second case will illustrate how alias frequencies in the overparameterized model can harm the model's ability to learn the target function. To demonstrate this, we now consider a noiseless, single-mode signal $g(x) := \hat{g}_{p} e^{i 2\pi p x}$ of frequency $p\leq n_0/2$. The data is hence of the form
\begin{align}\label{eq:target}
     y_j &= g(x_j) := \hat{g}_{p} e^{i 2\pi p x_j}.
\end{align}

Once again we choose $d = n(m+1)$ and for simplicity we assume an unweighted model, $\nu_k = 1$ for $k \in \Omega_d$. By orthonormality of Fourier basis functions, the interpolation condition requires that only the modes of the model $f$ with integer multiples of the frequency $p$ are retained. The interpolation constraint can then be rewritten as
\begin{equation}\label{eq:interp}
    \sum_{\ell=-m/2}^{m/2} \alpha_{p + n\ell}  = \hat{g}_{p}.
\end{equation}
The choice of \tcoef{s} $\alpha_k$ that satisfy Eq.~\eqref{eq:interp} while minimizing $\ell_2$-norm is
\begin{equation}\label{eq:zeronoise_sol}
    \alpha^{opt}_{p+n\ell} = \frac{\hat{g}_{p}}{m+1}
\end{equation}
for $k=p + n\ell$ and $\alpha_k=0$ otherwise (see Appendix~\ref{app:special_cases}). Eq.~\eqref{eq:interp} distributes the Fourier coefficient $\hat{g}_p$ among the \tcoef{s} $\alpha_{p + n\ell}$ corresponding to frequencies $p + n\ell$. Therefore, minimizing $\norm{\boldsymbol{\alpha}}_2$ in this case ``bleeds'' the target function into higher frequency aliases and results in the opposite effect compared to fitting a noisy signal (see Fig.~\ref{fig:noise_noiseless_demo}b): The generalization error of the overparameterized model now \textit{increases} with the number of aliases $m$ and the complexity of the model harms its ability to fit a noiseless target function. 

In order to recover a trade-off in generalization error for more general cases, we will need to consider more interesting distributions of \fweight{s} $\nu_k$ (instead of $\nu_k = 1$) that provide finer control over fitting the target function with low-frequency modes while spiking in the vicinity of noise with high-frequency aliases.

\subsection{Generalization trade-offs and benign overfitting}

The opposing effects of higher-frequency modes in overparameterized models in the cases discussed above hint at a trade-off in model performance that depends on the underlying signal and the \fweight{s} of the Fourier feature map. Returning to the more general case of input samples $y_j = g(x_j) + \epsilon_j$, in Appendix~\ref{app:classical_opt_err} we show that the task of fitting uniformly spaced samples using weighted Fourier features may be transformed into a linear regression problem, thereby generalizing the results of \cite{muthukumar2020harmless} to derive the following general solution to the minimum-$\ell_2$ interpolating problem of Eq.~\eqref{eq:classical_problem}:
\begin{equation}\label{eq:alpha_opt}
    \alpha^{opt}_{k + \ell n} =  \hat{y}_k \frac{\sqrt{\nu_{k + \ell n}}}{\sum_{j\in S(k)} \nu_{j}} ,
\end{equation}
where $\hat{y}_k$ is the discrete Fourier transform of $y_k$ and $k \in \Omega_{n_0}$, where 
\begin{equation}\label{eq:Sk}
S(k) = \{\ell: k + n\ell \in \Omega_d\}
\end{equation}
denotes the set of alias frequencies of $k$ appearing in the overparameterized model with spectrum $\Omega_d$. The optimal model is then expressed as
\begin{equation}\label{eq:opt_class}
    f^{opt}(x) = \sum_{k\in \Omega_{n_0}} \tilde{y}_k \sum_{\ell\in S(k)} e^{i2\pi \ell x} \frac{\nu_{\ell}}{\sum_{j\in S(k)} \nu_j}.
\end{equation}
Recalling that our model $f$ is trained on $n$ noisy samples  $(y_0, \dots, y_{n-1})$ of the target function $g$, we are interested in the squared error of the model $f$ averaged over (noisy) samples over the input domain, 
\begin{equation}\label{eq:gen_err}
    L(f) := \E_{x, \y} (f(x) - g(x))^2.
\end{equation}
and we call $L$ the \textit{generalization error} of $f$, as it captures the behavior of $f$ with respect to $g$ over the entire input domain $x \in[0, 1]$ instead of just the uniformly spaced training points $x_j$. In Appendix~\ref{app:classical_opt_err} we derive
\begin{align}
   L(f^{opt}) &= \underbrace{\frac{\sigma^2}{n} \sum_{k\in\Omega_n}   \frac{\sum_{\ell\in S(k)}\nu_\ell^2}{\left(\sum_{j\in S(k)} \nu_j \right)^2}}_\textsc{var}     + \underbrace{\sum_{k\in\Omega_{n_0}} |\hat{g}_k |^2 \left[ \left( \frac{\sum_{\ell\in S(k)\backslash k} \nu_\ell}{\sum_{j\in S(k)} \nu_j} \right)^2 + \frac{\sum_{\ell\in S(k)\backslash k} \nu_\ell^2 }{\left( \sum_{j\in S(k)} \nu_j\right)^2}\right]}_{\textsc{bias}^2} \label{eq:class_err}
\end{align}

We use this generalization error now to explore two interesting behaviors of the interpolating model in our setting: The tradeoff between noise absorption and signal recovery exemplified by the cases in Sec.~\ref{sec:cases}, and the ability of an overparameterized Fourier features model to \textit{benignly} overfit the training data.

The first behavior involves a trade-off in the \textit{generalization error} $L(f^{opt})$ between absorbing noise (reducing $\textsc{var}$) and capturing the target function signal (reducing $\textsc{bias}^2$) that recovers and generalizes the behavior of the two cases in Sec.~\ref{sec:cases}. This trade-off is controlled by three components: The noise variance $\sigma^2$, the input signal Fourier coefficients $\hat{g}_k$, and the distribution of \fweight{s} $\sqrt{\nu_k}$. As described in the two cases above, when $\sigma^2\rightarrow 0$ (signal only) the variance term $\textsc{var}$ vanishes and the model is biased for any choice of nonzero $\nu_{k}$ where $k>n$. Conversely, when $\hat{g}\rightarrow 0$ (noise only) the bias term $\textsc{bias}^2$ vanishes, and the variance term is minimized by choosing uniform $\nu_k$ for all $k\in\Omega_d$. 

The second interesting behavior occurs when the generalization error of the overparameterized model decreases at a nearly optimal rate as the number of samples $n$ increases, known as \textit{benign overfitting}. Prior work on benign overfitting in linear regression studied scenarios where the distribution of input data varied with the dimensionality of data and size of the training set in such a way that the excess generalization error of the overparameterized model (compared to a simple model) vanished \cite{bartlett2020benign}. However, since the dimensionality of the input data for our model is fixed, we instead consider sequences of \fweight{s} that vary with respect to the problem parameters ($n_0$, $n$, $d$) in a way that results in $\textsc{bias}^2$ and $\textsc{var}$ vanishing as $n\rightarrow \infty$. In this case, by fitting an increasing number of samples $n$ using such a sequence of \fweight{s}, the overparameterized model both perfectly fits the training data and generalizes well for unseen test points, and therefore exhibits benign overfitting.

\begin{figure}
    \centering
    \includegraphics[width=\textwidth]{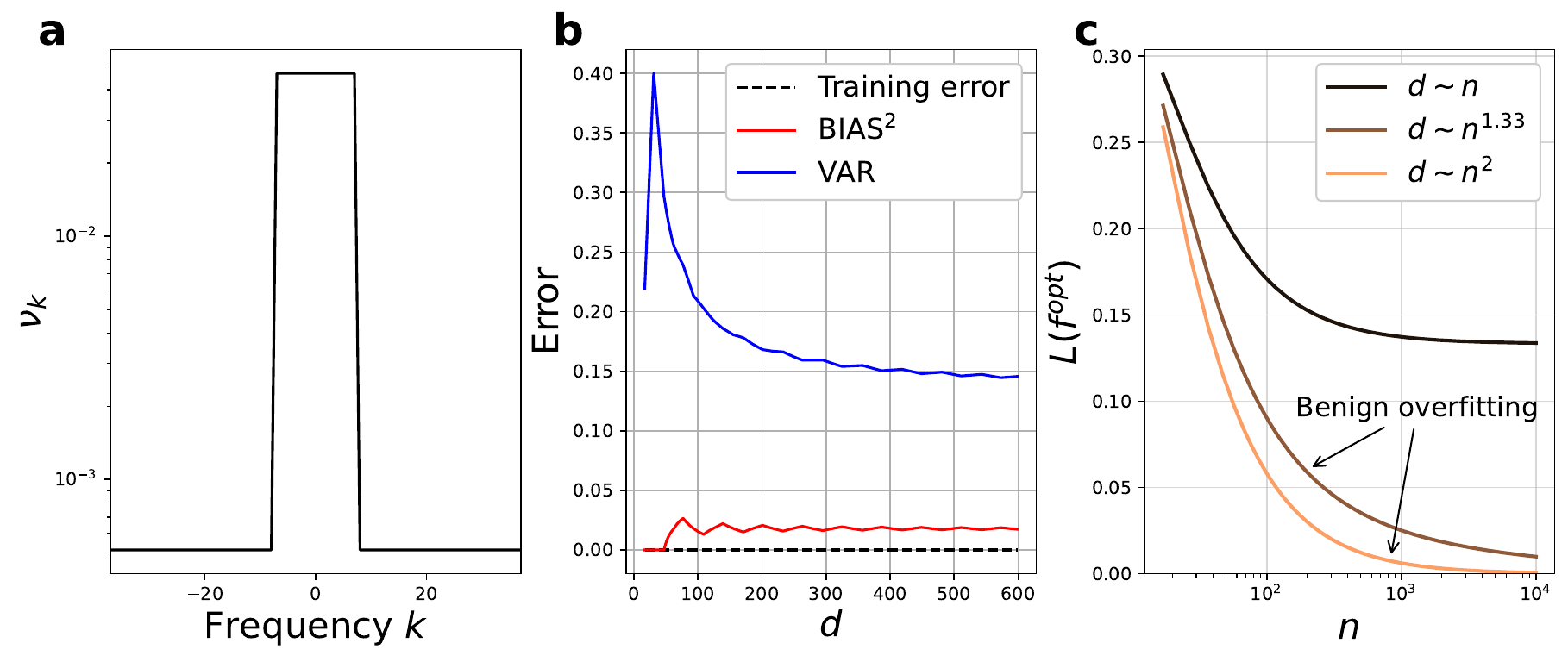}
    \caption{The \fweight{s} $\nu_k$ may be engineered to exhibit a bias-variance tradeoff and benign overfitting. \textbf{a} A demonstration of \fweight{s} $\nu_k$ which account for prior knowledge of the target function $g$, with large $\nu_k$ for $k \in \Omega_{n_0}$ and small $\nu_k$ for $k \in \Omega_d \backslash \Omega_{n_0}$. \textbf{b} After sampling a bandlimited target function ($n_0=15$) at a fixed number of uniformly spaced points ($n=31$), the generalization error of the minimum $\ell_2$-norm model $f^{opt}$ consists of a tradeoff between decreasing $\textsc{var}$ and increasing $\textsc{bias}^2$ for larger $d$. Note the peaking behavior for $\textsc{var}$ at $d/n = 1$. \textbf{c} This choice of \fweight{s} $\nu_k$ benignly overfits the input signal such that $\lim_{n,d\rightarrow \infty}L(f^{opt}) = 0$ for any scaling of $d =n^\alpha$ with $\alpha > 1$, and fails to generalize on the interval $[0, 1]$ when $\alpha \leq 1$.}
    \label{fig:hat_weights_bo}
\end{figure}

These behaviors are exemplified by a straightforward choice of \fweight{s} that incorporate some prior knowledge of the spectrum $\Omega_{n_0}$ available to the target function $g$. For all $k \in \Omega_{n_0}$, let $\nu_k = c/n_0$ for some positive $c$ and normalize the \fweight{s} so that $\sum_{k \in \Omega_k} \nu_k = 1$. We show in Appendix~\ref{sec:simple_bo} that the error terms of $L(f^{opt})$ scale as 
\begin{align}
    \textsc{var} &= \mathcal{O}\left( \frac{1}{n} + \frac{n}{d}\right), \label{eq:bound1} \\
    \textsc{bias}^2 &= \mathcal{O} \left( \frac{1}{n^2}\right).\label{eq:bound2}
\end{align}
Thus, as long as the dimension of the overparameterized Fourier features model grows strictly faster than $n$ (i.e., $d = \omega(n)$), the model
exhibits benign overfitting. In Appendix~\ref{sec:general_bo} we demonstrate how this simple example actually captures the mechanisms of benign overfitting for much more general choices of \fweight{s}. Fig.~\ref{fig:hat_weights_bo} summarizes this behavior and provides an example of the bias-variance tradeoff that occurs for overparameterized models. In particular,  Fig.~\ref{fig:hat_weights_bo}a exemplifies the setting in which benign overfitting occurs, wherein the \fweight{s} of the Fourier features model are strongly concentrated over frequencies in $\Omega_{n_0}$ but extend over a large range of alias frequencies for each $k \in \Omega_{n_0}$.

The generalization behavior described here is fundamentally different from many generalization guarantees typically found in statistical learning theory. While prior work has derived guarantees for the generalization of quantum models by constructing bounds on the complexity of the model class \cite{caro2021}, Eqs.~\ref{eq:bound1}-\ref{eq:bound2} demonstrate that generalization may occur as the complexity (i.e. dimension) of a model grows arbitrarily large.

So far, we have reviewed the Fourier perspective on fitting periodic functions in a classical setting and extended the analysis to characterize benign overfitting. However, if we can link the basic components of quantum models to the terms appearing in the error of Eq.~\eqref{eq:class_err}, then we will be able to study a similar trade-off in the error of overparameterized quantum models and the conditions necessary for benign overfitting. The remainder of this work is devoted to showing that analogous mechanisms exist in certain quantum machine learning models, and to studying the choices of \fweight{s} for which quantum models can exhibit tradeoffs in generalization error and benign overfitting.

\section{Benign overfitting in single-layer quantum models}\label{sec:qmodel}

In the previous section we have seen that the \fweight{s} $\sqrt{\nu_k}$ balance the trade-off between absorbing the noise and hurting the signal of overparametrized models. To understand how different design choices of quantum models impact this balance, we need to link their mathematical structure to the model class defined in Eq.~\eqref{eq:model}, and in particular to the \fweight{s}, which is what we do now.

The type of quantum models we consider here are simplified versions of parametrized quantum classifiers (also known as \textit{quantum neural networks}) that have been heavily investigated in recent years \cite{Mitarai_2018,neven_2018,Schuld_2020_centric}. They are represented by quantum circuits that consist of two steps: first, we encode a datapoint $x \in [0,1]$ into the state of the quantum system by applying a $\qdim$-dimensional unitary operator $V(x)$, and then we measure the expectation value of some $\qdim$-dimensional (Hermitian) observable $M$. This gives rise to a general class of quantum models of the form
\begin{equation}\label{eq:basic_qmodel}
    f(x) = \langle 0 | V^\dagger (x) M V(x) |0\rangle.
\end{equation}

To simplify the analysis, we will consider a quantum circuit $V(x)$ that consists of a data-independent unitary $U$ and a diagonal data-encoding unitary generated by a  $d$-dimensional Hermitian operator $H$,
\begin{align}
    S(x) = \exp(i 2\pi \cdot \text{diag}(\lambda_0, \dots, \lambda_{\qdim-1}) x) = \exp(i 2\pi H x),
\end{align}
which includes a large class of quantum models commonly studied but excludes schemes involving data re-uploading \cite{P_rez_Salinas_2020,jerbi2021}. Defining  $U\ket{0} = \ket{\Gamma} = \sum_{j=1}^\qdim \gamma_j \ket{j}$, the output of this quantum model becomes
\begin{align}
    f(x) = \bra{ \Gamma} S^\dagger(x) M S(x)|\Gamma\rangle,
\end{align}
where $|\Gamma\rangle$ can be treated as an arbitrary input quantum state. We call the corresponding quantum circuit for this model single-layer in the sense that it contains a single diagonal data-encoding layer in which all data-dependent circuit operations could theoretically be executed simultaneously (though in general the operation $U$ and measurement $M$ may require significant depth to implement).

Applying insights from Refs.~\cite{gil2020input,schuld2021}, quantum models of this form can be expressed in terms of a Fourier series 
\begin{align}\label{eq:gen_qmodel}
    f(x) &= \sum_{k \in \Omega} e^{i 2\pi k x } \sum_{\ell,m \in R(k)} \gamma_\ell \gamma_m^*  M_{m \ell },
\end{align}
where the \textit{spectrum} $\Omega$ as well as the \textit{partitions} $R(k)$ depend on the eigenspectrum $\lambda(H)$ of the data-encoding Hamiltonian $H$:
\begin{align}
    \Omega &= \{( \lambda_\ell - \lambda_m): \lambda_\ell, \lambda_m \in \lambda(H) \} \\
    R(k) &= \{(\ell, m): \lambda_\ell - \lambda_m = k; \,\, \lambda_\ell, \lambda_m \in \lambda(H)\} \label{eq:partition}.
\end{align}

Comparing Eq.~\eqref{eq:gen_qmodel} to Eq.~\eqref{eq:model} we see that that the quantum model may be expressed as a linear combination of weighted Fourier modes, but it is not yet clear how the input state $\gamma_j$ and the trainable observable $M$ of the quantum model correspond to \fweight{s} $\nu_k$ for each Fourier mode. To reveal this correspondence, we will need to first find the minimum-norm interpolating observable that solves the optimization problem
\begin{align}\label{eq:qproblem} 
    M^{opt} = \argmin_{M = M^\dagger: f(x_j) = y_j \,\forall\, j} \norm{M}_F ,
\end{align}
where $\norm{M}_F = \sqrt{\tr{M^\dagger M}}$ denotes the Frobenius norm of $M$. Solving Eq.~\eqref{eq:qproblem} is analogous to the minimization the $\ell_2$ norm of $\balpha$ in the classical optimization problem of Eq.~\eqref{eq:classical_problem}, and serves a role similar to regularization commonly applied to quantum models by introducing a penalty term proportional to $\norm{M}_F^2$ \cite{huang2021power,gyurik_structural_2021,schuld_supervised}. In Appendix~\ref{app:quantum_opt_err} we prove that subject to the condition that $\gamma_i > 0$, the minimum-$\norm{\cdot}_F$ interpolating observable that solves Eq.~\eqref{eq:qproblem} is given as
\begin{equation}
    (M^{opt})_{m\ell}  =  \hat{y}_k  \frac{\gamma_\ell^* \gamma_m}{\sum_{i,j \in R(S_k)} |\gamma_i|^2 |\gamma_j|^2},
\end{equation}
for all $\ell, m \in R(S_k)$, and the corresponding optimal quantum model  is

\begin{equation}\label{eq:gen_qmodel_opt}
f^{opt}(x) =  \sum_{k\in \Omega_{n_0}} \hat{y}_k \, \sum_{\ell\in S(k)}  e^{i 2\pi \ell x} \frac{\nu^{opt}_\ell}{{\sum_{j \in S(k)} \nu^{opt}_j}},
\end{equation}
where $S(k)$ denotes the set of aliases of $k$ appearing in $\Omega$ from Eq.~\eqref{eq:Sk}. By comparison to the optimal classical model of Eq.~\eqref{eq:opt_class} we have identified the \fweight{s} of the optimized quantum model as
\begin{equation}\label{eq:nu_comp}
    \nu_{k}^{opt} := \sum_{i, j \in R(k)} |\gamma_i|^2 |\gamma_j|^2.
\end{equation}

Interestingly, while there was initially no clear way to separate the building blocks of the quantum model in Eq.~\eqref{eq:gen_qmodel} into \tcoef{s} $\alpha_k$ and \fweight{s} $\nu_k$, this separation has now appeared after solving for the optimal observable $M^{opt}$. Furthermore, the optimal quantum model depends on $|\gamma_i|$ and is independent of phases associated with amplitudes $\gamma_i$ (an effect that stems from using only a single data-encoding layer $S(x)$ in the quantum model).

From Eq.~\eqref{eq:nu_comp} it is clear that the partitions $R(k)$ of Eq.~\eqref{eq:partition} arising from the choice of data-encoding unitary $S(x)$ have a strong relationship with the \fweight{s} $\nu_k$ of the quantum model. We will now consider a simplified quantum model to highlight this relationship, thereby identifying a tradeoff between noise absorption and target signal recovery and the possibility of observing benign overfitting in quantum models.

\subsection{Simplified quantum model}\label{sec:toy_model}

To explicitly highlight the role of $R(k)$ in controlling the \fweight{s} $\nu_k^{opt}$ of the optimized quantum model, we will now simplify the model by using an equal superposition input state $|\Gamma\rangle = \frac{1}{\sqrt{d} }\sum_{j=0}^{d-1} \ket{j}$ and by restricting the set of observables considered during optimization. If we fix every entry of the observable with respect to elements in a partition $R(k)$ to be proportional to some complex constant $M(k)$: 
\begin{equation}\label{eq:restrictedM}
M_{m \ell} = \frac{M(k)}{\sqrt{|R(k)|}} \qquad \text{ for all } \ell, m \in R(k),
\end{equation}
then we can simplify the quantum model of Eq.~\eqref{eq:gen_qmodel} to
\begin{align}\label{eq:toy_qmodel}
    f(x) &= \sum_{k \in \Omega} M(k) \frac{ \sqrt{|R(k)|}}{d}e^{i 2\pi k x }.
\end{align}

Comparing Eq.~\eqref{eq:toy_qmodel} to Eq.~\eqref{eq:model} we identify a direct correspondence between the \tcoef{s} $\alpha_k$ in the classical model with $M(k)$, as well as a correspondence between the \fweight{s} $\sqrt{\nu_k}$ and the the \textit{degeneracy} $|R(k)|$ of the quantum model. Making the substitutions $\alpha_k \rightarrow M(k)$, $\sqrt{\nu_k} \rightarrow \sqrt{|R(k)|}/d$, one can verify that $\sum_{k\in \Omega} |M(k)|^2 = \norm{M}_F^2$ for this restricted choice of $M$ and so the solution to the optimization of Eq.~\eqref{eq:qproblem} is essentially the same as that of the classical problem in Eq.~\eqref{eq:alpha_opt}. 

The crucial property of the simplified model is that the degeneracy $|R(k)|$ -- and hence the combinatorial structure introduced by the data encoding Hamiltonian's eigenvalues -- completely controls the trade-off in the generalization error (Eq.~\eqref{eq:class_err}). We can hence study different types of partitions $R(k)$ to show a direct effect of the data-encoding unitary $S(x)$ on the fitting and generalization error behaviors for this simplified, overparameterized quantum model. 

To study these behaviors we will now consider specific families of $H$ which we call \textit{encoding strategies} since the choice of $H$ completely determines how the data is encoded into the quantum model. While $R(k)$ and $\Omega$ may be computed for an arbitrary $S(x)$ using brute-force combinatorics, some encoding strategies lead to particularly simple solutions. We have derived a few such examples of simple degeneracies and spectra for different encoding strategies in  Appendix~\ref{app:combinatorics} and present the results in Table~\ref{tab:degeneracy_sets}. These choices highlight the extreme variation in $\Omega$ resulting from minor changes to $S(x)$, for example $|\Omega| \propto n_q$ for the ``Hamming'' encoding strategy compared to $|\Omega| \propto 3^{n_q}$ for the ``Ternary'' encoding strategy. These examples also highlight the limitations in constructing Hamiltonians with specific properties such as uniform $|R(k)|$ or evenly-spaced frequencies in $\Omega$.

\begin{table}
    \centering
    \begin{tabular}{|l|l|l|l|l|}
    \hline
        \makecell{Encoding \\strategy} &  Hamiltonian example & \makecell{Degeneracy \\$|R(k)|$}  & Spectrum $\Omega$ & $|\Omega|$ \\
        \hline
       Hamming  & $\frac{1}{2}(Z_0 + Z_1 + Z_2 + \dots)$ & $\begin{pmatrix} 
       2n_q \\ n_q-|k|
       \end{pmatrix}$ & $ \{ -n_q, 1-n_q, \dots, n_q\}$ & $ 2n_q + 1$  \\
       Binary  & 
       $\begin{aligned}
       \textstyle&\frac{1}{2}(Z_0 + 2Z_1 + 4Z_2 + \dots) \\
       &\sim \diag{0, 1, 2, \dots, 2^{n_q}-1}
       \end{aligned} $
       & $ 2^{n_q} - |k|$& $\{ -2^{n_q}+1,  \dots 2^{n_q} - 1\}$ & $2^{n_q+1} - 1$  \\
       Ternary  & $\frac{1}{2}(Z_0 + 3Z_1 + 9Z_2 + \dots)$ & $2^{n_q - \norm{T(k)}_1}$ & $\Bigl\{- \Bigl\lfloor \frac{3^{n_q}}{2}\Bigr\rfloor , \dots,  \Bigl\lfloor \frac{3^{n_q}}{2} \Bigr\rfloor \Bigr\}$& $3^{n_q}$  \\
        Golomb  & $\diag{0, 1, 4, 6}$ & $\begin{cases} d & k=0 \\ 1 & k\neq 0
       \end{cases}$ & varies & $2\begin{pmatrix} 
       d \\ 2
       \end{pmatrix}+ 1$  \\
       \hline
    \end{tabular}
    \caption{ Spectra $\Omega$ and degeneracies $R(k)$ computed for various data-encoding Hamiltonians $H$ defined for either $d$ dimensions or $n_q$ qubits. The Hamming, Binary, and Ternary data encoding strategies are realized on $n_q$ qubits using a separable  Hamiltonian consisting of Pauli-$Z$ operators with different prefactor schemes. The Ternary encoding strategy results in the largest $|\Omega|$ possible for a separable data encoding Hamiltonian (see also Ref.~\cite{shin2022}), while the Golomb encoding encoding strategy (named in reference to Golomb rulers, e.g.~\cite{piccard1939}) results in the largest $|\Omega|$ possible for any choice of $d$-dimensional data-encoding Hamiltonian. Note that the spectrum is preserved under permutations and additive shifts of the diagonal of the Hamiltonian and so we use ``$\sim$'' to denote equivalence up to these operations. The function $T$ converts an integer to a signed ternary string, as defined in Eq.~\eqref{eq:shift_ternary} (see Appendix~\ref{app:combinatorics}). }
    \label{tab:degeneracy_sets}
\end{table}

\begin{figure}
    \centering
    \includegraphics[width=\textwidth]{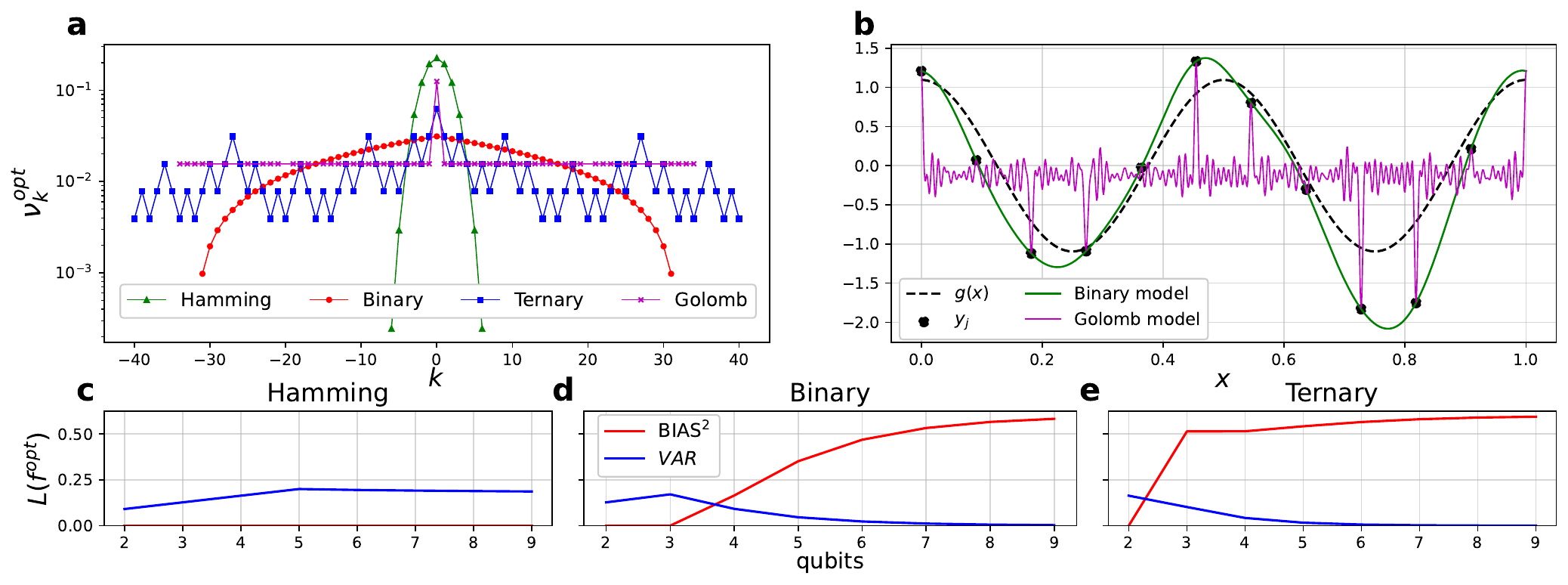}
    \caption{The degeneracies $|R(k)|$ for select quantum models directly influence the tendency of the models to overfit noise or bleed the underlying signal. \textbf{a} The degeneracies $|R(k)|$ for each frequency $k$ for the four models introduced in Table~\ref{tab:degeneracy_sets} induce significantly different \fweight{s} in the simplified quantum model. \textbf{b} Two models exemplify the extreme cases of fitting behaviors from Sec.~\ref{sec:cases}: The Golomb encoding strategy (magenta) has significant weight on \fweight{s} $\nu_k$ for large frequencies $k$, resulting in a ``spiky'' overfitting behavior that interpolates each noisy sample $y_j$ but sacrifices recovery of the underlying signal $g(x)$; the Hamming encoding strategy (green) fits noisy samples smoothly and the error of the model is sensitive to the noise in the sampled data. \textbf{c}-\textbf{e} The $\textsc{Bias}^2$ and $\textsc{var}$ of the optimized  Hamming encoding strategy (\textbf{c}), Binary encoding strategy (\textbf{d}), and Ternary encoding strategy (\textbf{e}) demonstrate the general effect of $|R(k)|$ and $\nu_k$ on the generalization error. The concentration of \fweight{s} $\nu_k$ on $k \in \Omega_{n_0}$ in the Hamming encoding strategy reduces the $\textsc{bias}$ (\textbf{c}), while $\nu_k$ having broad support over $k\in\Omega$ (e.g. Ternary, \textbf{e}) reduces $\textsc{var}$ as the resulting $f$ is close to its average (with respect to noise) except for becoming ``spiky'' in the vicinity of samples $y_j$, but increases the $\textsc{bias}$ by introducing alias frequencies into the optimized model $f^{opt}$.}
    \label{fig:simple_model_fits}
\end{figure}

Since the \fweight{s} $\nu_k$ of the Fourier modes are fixed by the choice of the data-encoding unitary, we can understand a choice of $S(x)$ as providing a structural bias of a quantum model towards different overfitting behaviors, and conversely the choices of \fweight{s} available to quantum models are limited and are particular to the structure of the associated quantum circuit. Figure~\ref{fig:simple_model_fits} shows distributions for \fweight{s} arising from the example encoding strategies presented in Table~\ref{tab:degen_tedious}, and demonstrates a broad connection between the degeneracies $|R(k)|$ of the model (giving rise to \fweight{s} $\nu_k^{opt}$) and the generalization error $L(f^{opt})$.

\newpage
\subsection{Trainable unitaries reweight the general quantum model}\label{sec:qmodel2}

We now return to the general quantum model of Eq.~\eqref{eq:gen_qmodel} to understand  how different choices of the state preparation unitary $U$ (giving rise to input state $|\Gamma\rangle$) affect the \fweight{s} $\nu_k^{opt}$ of the general quantum model, thereby influencing benign overfitting of the target function. While Eq.~\eqref{eq:nu_comp} differs from the correspondence $\nu_k = |R(k)| / d^2$ that we observed for the simplified quantum model, we see that the \fweight{s} of the optimal quantum model still depend heavily on the degeneracies $|R(k)|$. For instance, the \textit{average} $\nu_k^{opt}$ with respect to Haar-random $d$-dimensional input states $|\Gamma\rangle$ is proportional to $|R(k)|$ whenever $k \neq 0$:
\begin{align}
   \underset{U \sim \U{d}}{\E} \nu_k^{opt}  = \frac{|R(k)| + d \delta_0^k}{d(d+1)} .
\end{align}
where $\delta_i^j$ denotes the Kronecker delta. Furthermore, in Appendix~\ref{app:quantum_opt_err} we observe that the variance of $\nu_k^{opt}$ around its average tends to be small for encoding strategies considered in this work, for instance scaling like $\mathcal{O}\left(d^{-3}\right)$ for the Binary encoding strategy and $\mathcal{O}\left(d^{-4}\right)$ for the Golomb encoding strategy. This demonstrates that the \fweight{s} of generic quantum models (i.e., ones for which $U$ is randomly sampled) will be dominated by the degeneracy $|R(k)|$ introduced by the data-encoding operation $S(x)$. 

Despite the behavior of $\nu_k$ being dominated by $|R(k)|$ in an average sense, there are specific choices of $U$ for which the \fweight{s} deviate significantly from this average. We will now use one such choice of $U$ to provide a concrete example of an interpolating quantum model that exhibits benign overfitting. Suppose we choose $U$ such that the elements of $|\Gamma\rangle$ are given by
\begin{align}\label{eq:gamma_hat_main}
    \gamma_j = \begin{cases}
    \sqrt{a}, & j \in [c_1, c_2), \\
    \sqrt{b}, & \text{otherwise}
    \end{cases}
\end{align}
for $j\in [d]$, and some integers $0 < c_1 < c_2 < d$ and amplitudes $a, b$ subject to normalization. We show that given a band-limited target function $g$ with access to spectrum $\Omega_{n_0}$, there is a specific choice of $c_1, c_2$ dependent on $d$ and $n_0$ for which the interpolating quantum model $f^{opt}$ of Eq.~\eqref{eq:gen_qmodel_opt} also benefits from vanishing generalization error in the limit of many samples, namely we show in Appendix~\ref{app:quantum_bo} that
\begin{equation}
    L(f^{opt}) = \mathcal{O}\left( \frac{1}{n} + \frac{n}{d} \right).
\end{equation}

\begin{figure}
    \centering
    \includegraphics[width=\textwidth]{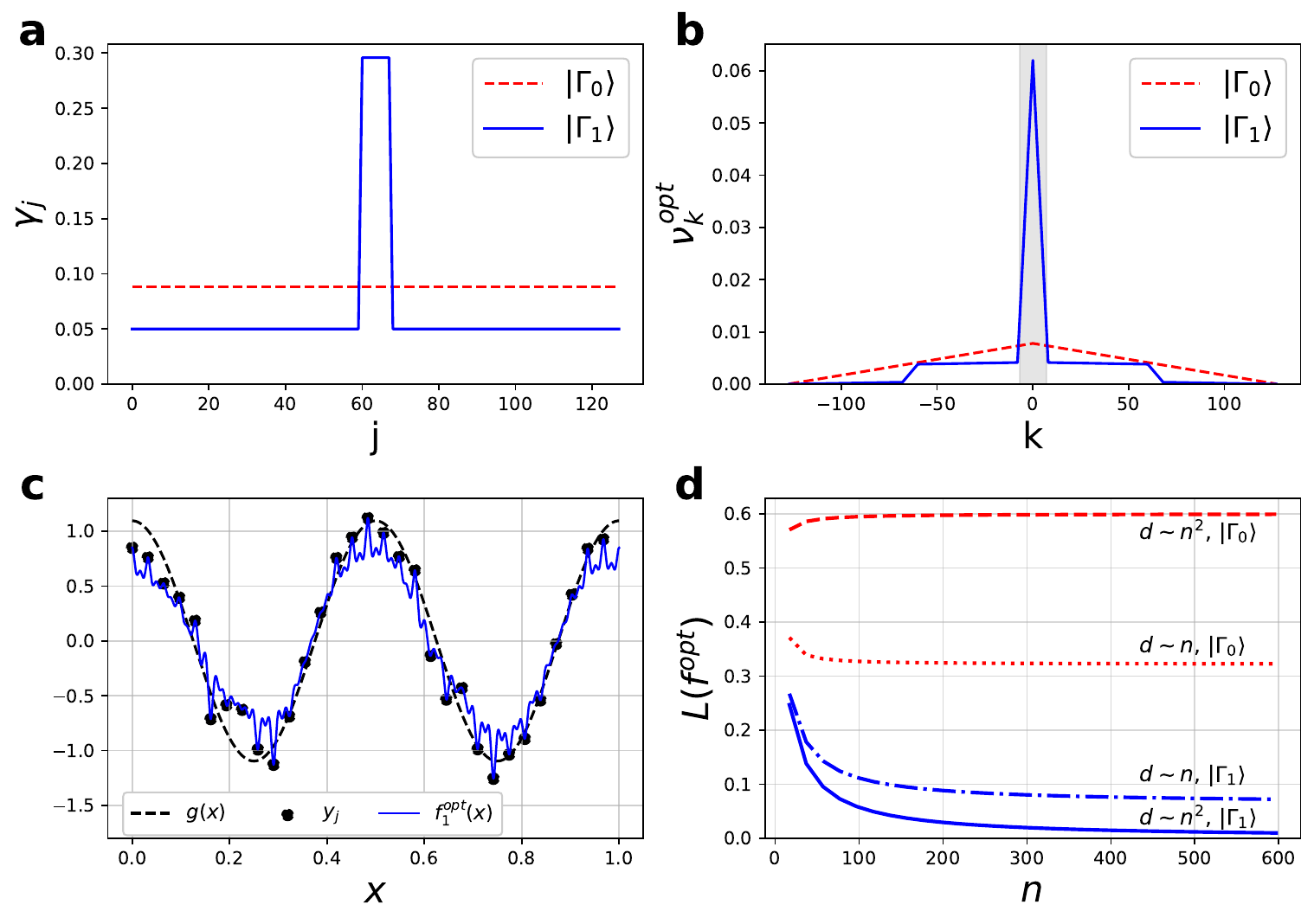}
    \caption{Benign overfitting for the general quantum model. \textbf{a} For the Hamiltonian $H \sim \diag{(0, 1, \dots, d-1)}$ we construct a simple input state $|\Gamma_1\rangle$ (Eq.~\eqref{eq:gamma_hat_main}) that is dependent on $n_0$ and $d$ (see Appendix~\ref{app:quantum_bo}), compared to the uniform state $|\Gamma_0\rangle = \frac{1}{\sqrt{d}} \sum_k |k\rangle$. \textbf{b} $|\Gamma_1\rangle$ induces \fweight{s} $\nu_k^{opt}$ in the optimized model that are heavily concentrated in the target function spectrum $\Omega_{n_0}$ (gray band) and decay like $1/d$ and $1/d^2$ elsewhere for $k \in \Omega_d \backslash \Omega_{n_0}$, compared to the \fweight{s} induced by $|\Gamma_0\rangle$ which have similar magnitude over all of $\Omega_d$. \textbf{c} Plotting the $\norm{M}_F$-minimizing quantum model $f_1^{opt}$ with input state $|\Gamma_1\rangle$ for a sample signal ($d=128$, $n=31$) highlights how $|\Gamma_1\rangle$ and $H$ work together to interpolate $y_j$ via high-frequency ``spiky'' behavior near the sampled points, but otherwise closely matches the underlying signal $g(x)$. \textbf{d} The resulting quantum model benignly overfits the band-limited target function: As long as $d = \omega(n)$, the generalization error  $L(f_1^{opt})$ of the minimum-$\norm{M}_F$ interpolating quantum model vanishes for large enough sample size $n$.}
    \label{fig:quantum_bo}
\end{figure}

Thus, by perfectly fitting the training data and exhibiting good generalization in the $n\rightarrow \infty$ limit, the quantum model exhibits benign overfitting. This behavior is outlined in Figure~\ref{fig:quantum_bo}, which highlights the role that $|\Gamma\rangle$ plays in concentrating the \fweight{s} $\nu_k$ within the spectrum $\Omega_{n_0}$ of $g$ while preserving a long tail that provides the model with low-variance ``spiky'' behavior in the vicinity of noisy samples. In contrast, the \fweight{s} for the Binary encoding strategy with a uniform input state has little support on $\Omega_{n_0}$ and resulting in a large $\textsc{bias}$ error.

The above discussion shows how the input state amplitudes $\gamma_i$ provide additional degrees of freedom with which the \fweight{s} $\nu_k^{opt}$ can be tuned in order to modify the generalization behavior of the interpolating quantum model, and to exhibit benign overfitting in particular. It is therefore worthwhile to consider what other kinds of \fweight{s} $\nu_k$ might be prepared by some choice of input state $|\Gamma\rangle$. We may use simple counting arguments to demonstrate the restrictions in designing particular distributions of \fweight{s}. Suppose we define $\Omega_+ = \{ k: k \in \Omega, k>0\}$ containing the positive frequencies of a quantum model. Then the introduction of an arbitrary input state $|\Gamma\rangle$ provides us with $2^{n_q} - 1$ free parameters with which to tune $|\Omega_+|$-many terms in the distribution of $\nu_k^{opt}$ (subject to $\nu_k = \nu_{-k}$ and $\sum_k \nu_k = 1$). Clearly, there are distributions of \fweight{s} $\nu_k^{opt}$ that can not be achieved for models where $|\Omega_+| \geq 2^{n_q}$. 

Conversely, the condition $|\Omega_+| < 2^{n_q}$ does not necessarily mean that we can thoroughly explore the space of possible \fweight{s} by modifying the input state $|\Gamma\rangle$. For example, consider the Hamming encoding strategy for which the number of free parameters controlling the distribution of \fweight{s} $\nu_k^{opt}$ is $|\Omega_+| = n_q$, which is exponentially smaller  the number of parameters in $|\Gamma\rangle$. While this might suggest significant freedom in adjusting $\nu_k^{opt}$, the opposite is true: For any choice of input state $|\Gamma\rangle$, there is another state of the form
\begin{equation}\label{eq:hamming_invariance}
    |\Gamma'\rangle = \sum_{i=0}^{n_q} \phi_i |\Phi_i\rangle,
\end{equation}
that achieves exactly the same distribution of \fweight{s} $\nu_k$. In Eq.~\eqref{eq:hamming_invariance}, $|\Phi_i\rangle$ describes a uniform superposition over all computational basis state bitstrings with weight $i$, and so the distribution of $\nu_k^{opt}$ actually only depends on $n_q + 1$ real parameters $\phi_i$, $i=0, 1, \dots, n_q$, and the \fweight{s} are invariant under any operations in $U$ that preserve $|\Phi_i\rangle$ (see Appendix~\ref{app:quantum_opt_err}). An example of such operations are the \textit{particle-conserving unitaries} well-known in quantum chemistry, which act to preserve the number of set bits (i.e., the Hamming weight) when each bit represents a mode occupied by a particle in second quantization \cite{PhysRevA.92.062318,PhysRevLett.120.110501}. This example demonstrates how symmetry in the data-encoding Hamiltonian (e.g. Refs.~\cite{larocca2022,meyer2022}) can have a profound influence on the ability to prepare specific distributions of \fweight{s} $\nu_k^{opt}$, and consequently affect the generalization and overparameterization behavior of the associated quantum models.



\section{Conclusion}

In this work we have taken a first step towards characterizing the phenomenon of benign overfitting in a quantum machine learning setting. We derived the error for an overparameterized Fourier features model that interpolates the (noisy) input signal with minimum $\ell_2$-norm \tcoef{s} and connected the \fweight{s} associated with each Fourier mode to a trade-off in the generalization error of the model. We then demonstrated an analogous simplified quantum model for which the \fweight{s} are induced by the choice of data-encoding unitary $S(x)$. Finally, we discussed how introducing an arbitrary state-preparation unitary $U$ gives rise to effective \fweight{s} in the optimized general quantum model, presenting the possibility of connecting $U$ and $S(x)$ to benign overfitting in more general quantum models. 

Our discussion of interpolating quantum models presents an interpretation of overparameterization (i.e., the size of the model spectrum $\Omega$) that departs from other measures of quantum circuit complexity discussed in the literature \cite{Abbas_2021,larocca2021theory,Du_2020}, as even the simplified quantum models studied here are able to interpolate training data using a fixed circuit $U$ and optimized measurements. We also reemphasize that -- unlike much of the quantum machine learning literature -- we do not consider a setting where the model is optimized with respect to a trainable circuit, as the model of Eq.~\eqref{eq:gen_qmodel_opt} is constructed to exhibit zero training error (and can therefore not be improved via optimization). Finding the input state $|\Gamma\rangle$ that will result in a specific distribution of \fweight{s} $\nu_k^{opt}$ generally requires solving a $|\Omega_+|$-dimensional system of equations that are second order in $2^{n_q}$ many real parameters $|\gamma_i|^2$ (i.e., inverting the map of the form $\mathbb{R}^{2^{n_q}} \rightarrow \mathbb{R}^{|\Omega_+|}$ in Eq.~\eqref{eq:nu_comp}) or otherwise performing a variational optimization that will likely fail due to the familiar phenomenon of \textit{barren plateaus} \cite{mcclean2018barren,cerezo2021cost,holmes2022connecting,wang2021noise}.

While we have shown an example of benign overfitting by a quantum model in a relatively restricted context, future work may lead to more general characterizations of this phenomenon. Similar behavior likely exists for quantum kernel methods and may complement existing studies on these methods' generalization power \cite{canatar2022}. An exciting possibility would be to demonstrate benign overfitting in quantum models trained on distributions of quantum states which are hard to learn classically \cite{huang2021experiment,chen2022}, thereby extending the growing body of statistical learning theory for quantum learning algorithms \cite{caro2021generalization,PRXQuantum.2.040321,huang2021inform}.


\section{Code availability}

Code to reproduce the figures and analysis is available at the following repository: \url{https://github.com/peterse/benign-overfitting-quantum}.

\section{Acknowledgements}
The authors thank Nathan Killoran, Achim Kempf, Angus Lowe, and Joseph Bowles for helpful feedback. This work was supported by Mitacs through the Mitacs Accelerate program. Research at Perimeter Institute is supported in part by the Government of Canada through the Department of Innovation, Science and Economic Development and by the Province of Ontario through the Ministry of Colleges and Universities. Circuit simulations were performed in PennyLane \cite{pennylane}. 

\bibliographystyle{quantum}

\onecolumn
\appendix

\section{Solution for the classical overparameterized model}\label{app:classical_opt_err}

In this section we derive the optimal solution and generalization error for the classical overparameterized  weighted Fourier functions model. We then discuss the conditions under which benign overfitting may be observed and construct examples of the phenomenon.

\subsection{Linearization of overparameterized Fourier model}

We first show that the classical overparameterized Fourier features model may be cast as a \textit{linear} model under an appropriate orthogonal transformation. We are interested in learning a target function of the form

\begin{equation}\label{eq:fdeffourier}
    g(x) = \sum_{k \in \Omega_{n_0}} \hat{g}_k e^{i 2\pi k x},
\end{equation}
with the additional constraint that the Fourier coefficients satisfy $\hat{g}_k = \hat{g}_{-k}^*$ such that $g$ is real. The spectrum of Eq.~\eqref{eq:fdeffourier} for odd $n_0$ only contains integer frequencies,
\begin{equation}\label{eq:class_spec}
   \Omega_{n_0} = \Biggl\{-\frac{n_0 - 1}{2}, \dots, 0, \dots, \frac{n_0 - 1}{2} \Biggr\},
\end{equation} 
and we accordingly call $g$ a \textit{bandlimited} function with bandlimit $n_0/2 - 1$. To learn $g$, we first sample $n$ equally spaced datapoints on the interval $[0, 1]$,
\begin{equation}\label{eq:uniformx}
    x_j = \frac{j}{n}, \qquad j\in [n],
\end{equation}
where $[n] = \{0, 1, 2, \dots, n-1\}$ and we assume $n$ is odd, and we then evaluate $g(x_j)$ with additive error. This noisy sampling process yields $n$ observations of the form $y_j = g(x_j) + \epsilon$ with $\E[\epsilon^2] = \sigma^2$. We will fit the observations $y_j$ using overparameterized Fourier features models of the form
\begin{equation}\label{eq:fff}
    f(x) = \langle \phi(x), \balpha \rangle = \sum_{k \in \Omega_d} \alpha_k \sqrt{\nu_k} e^{i 2\pi k x},
\end{equation}
with $\balpha \in \C^d$, and we have introduced weighted Fourier features $\phi: \R \rightarrow \C^d$ defined elementwise as
\begin{equation}
    \phi(x)_k = \sqrt{\nu_k} e^{i 2 \pi k x}.
\end{equation}

In Eq.~\eqref{eq:fff}, $\Omega_d$ describes the set of frequencies available to the model for any choice of $d \geq n \geq n_0$. We are interested in the case where $f$ interpolates the observations $y_j$, i.e., $f(x_j) = y_j$ for all $j=0, \dots, n-1$. To this end, we define a $n\times d$ feature matrix $\Phi$ whose rows are given by $\phi(x_j)^\dagger$:
\begin{align}
    \Phi_{jk} = [\phi(x_j)]_k^* = \sqrt{\nu_k} e^{-i 2\pi j k / n}.
\end{align}
The interpolation condition may then be stated in matrix form as 
\begin{equation}\label{eq:interp_matrix}
    \Phi \balpha = \y,
\end{equation}
where $(\y)_j = y_j$ is the vector of noisy observations. $\Omega_d$ contains alias frequencies of $\Omega_{n_0}$, and so there the choice of $\balpha$ that satisfies Eq.~\eqref{eq:interp_matrix} is not unique. Here we will focus on the minimum-$\ell_2$ norm interpolating solution,
\begin{equation}\label{eq:alpha_opt_prob}
    \balpha^{opt} =  \argmin_{\Phi \balpha = \y} \norm{\balpha}_2.
\end{equation}

\subsubsection{Fourier transform into the linear model}

We will now show that Eq.~\eqref{eq:alpha_opt_prob} with uniformly-spaced datapoints $x_j$ can be solved using methods from ordinary linear regression under a suitable choice of transformation. Defining the $n$-th root of identity as $\omega = e^{i 2\pi / n}$, then the $j$-th row of the LHS of Eq.~\eqref{eq:interp_matrix} is equivalent to 
\begin{align}
    \sum_{k\in\Omega_d} \alpha_k \sqrt{\nu_k} \omega^{-kj} &=  \sum_{k\in\Omega_n} \sum_{\ell \in S(k)} \alpha_{k + n\ell} \sqrt{\nu_{k + n\ell}} \omega^{-(k + n\ell)j}
    \\&= \sum_{k\in\Omega_n} \omega^{-kj} \sum_{\ell \in S(k)} \alpha_{k + n\ell} \sqrt{\nu_{k + n\ell}} 
    \\&:= \sum_{k \in \Omega_n} \omega^{-kj} \langle P_{S(k)} \balpha, P_{S(k)} \mathbf{w}\rangle.
\end{align}
where $S(k) = \{ j: j \in \Omega_d, j \text{ mod } n = k \}$ is the set of alias frequencies of $k$ appearing in $\Omega_d$, i.e. the set of frequencies $k + \ell n$ with $\ell \in \Z$ that obey
\begin{equation}
    e^{i2\pi k x_j} = e^{i2\pi k j/n} = e^{i2\pi (k + \ell n)j/n} = e^{i2\pi (k + \ell n)x_j},
\end{equation}
which follows from $e^{i 2\pi \ell k} = 1$ since $k \in \Z$. The operator $P_{S(k)}: \C^d \rightarrow \C^d$ is the projector onto the set of standard basis vectors in $\C^d$ with indices in $S(k)$, and $(\mathbf{w})_k := \sqrt{\nu_k}$. The roots of unity are orthonormal since
\begin{align}
    \sum_{k = 0}^{n-1} \omega^{k (p-q)} = n \delta_p^q
\end{align}
for $p, q \in [n]$. This implies
\begin{align}\label{eq:ident_omega}
 \sum_{k \in \Omega_n} \omega^{k (p-q)} = \omega^{\min(\Omega_n)(p-q)} n \delta_p^q = n\delta_p^q,
\end{align}
for $p, q \in \Omega_n$. Defining the discrete Fourier Transform $\hat{y}_k$ of $\y$ according to
\begin{equation}\label{eq:dft}
    \hat{y}_j = \frac{1}{n}\sum_{k \in \Omega_n} y_k \omega^{jk},
\end{equation}
then using the identity of Eq.~\eqref{eq:ident_omega}, we evaluate the $j$-th row of Eq.~\eqref{eq:interp_matrix} as
\begin{align}
    \sum_{k \in \Omega_n} \omega^{-kj} \langle P_{S(k)} \balpha, P_{S(k)} \mathbf{w}\rangle &= y_j,
    \\ \sum_{j \in \Omega_n} \omega^{p j}\sum_{k \in \Omega_n} \omega^{-kj} \langle P_{S(k)} \balpha, P_{S(k)} \mathbf{w}\rangle &= \sum_{j \in \Omega_n} \omega^{p j} \left(\sum_{k \in \Omega_n} \hat{y}_k \omega^{- kj} \right),
    \\ \sum_{k \in \Omega_n} n \delta_k^p \langle P_{S(k)} \balpha, P_{S(k)} \mathbf{w}\rangle &=  \sum_{k \in \Omega_n} n \delta_k^p \hat{y}_k,
    \\ \langle P_{S(p)} \balpha, P_{S(p)} \mathbf{w}\rangle &= \hat{y}_p. \label{eq:final_fourier}
\end{align}

Inspecting this final line yields a new matrix equation: Let $X$ be an $n\times d$ matrix $X$ with elements
\begin{align}\label{eq:Xjk}
    X_{jk} = (P_{S(j)} \mathbf{w})_k = \sqrt{\nu_k} \cond{k \in S(j)},
\end{align}
where the conditional operator $\cond{Z}$ evaluates to $1$ if the predicate $Z$ is true, and $0$ otherwise. Then we may express Eq.~\eqref{eq:final_fourier} for all $p \in \Omega_n$ as a matrix equation
\begin{equation}\label{eq:interp_matrix_fourier}
    X \balpha = \hat{\y},
\end{equation}
where $(\hat{\y})_j = \hat{y}_j$. We have shown that Eq.~\eqref{eq:interp_matrix} is exactly equivalent to Eq.~\eqref{eq:interp_matrix_fourier} for uniformly spaced inputs, and as $\balpha$ is unchanged between these two representations this implies that the solution to Eq.~\eqref{eq:alpha_opt_prob} is also given by
\begin{equation}\label{eq:alpha_opt_prob_fourier}
    \balpha^{opt} = \argmin_{X \balpha = \hat{\y}} \norm{\balpha}_2.
\end{equation}

Therefore, the minimum $\ell_2$-norm solution to interpolate the input signal using weighted Fourier functions provided as samples $y_j$ is exactly the same as the minimum $\ell_2$-norm solution for an equivalent \textit{linear regression} problem on the matrix $X$ with targets $\hat{\y}$. Furthermore, this linear regression problem is related to the original problem via Fourier transform. Let $F$ be the (nonunitary) discrete Fourier transform defined on $\C^n$ elementwise as 
\begin{align}
    F_{jk} &= \omega^{jk}, \\
    F^\dagger F &= n\I_n.
\end{align}
Then for $k \in \Omega_d$, $j \in \Omega_n$,
\begin{align}
\frac{1}{n}(F \Phi)_{jk} = \frac{1}{n} \sum_{\ell \in \Omega_n}  \omega^{(k-j)\ell}  \sqrt{\nu_k} = \frac{1}{n} \left(n \delta_j^{k \text{ mod }  n}\right) \sqrt{\nu_k} = \sqrt{\nu_k} \cond{k \in S(j)} = X_{jk},
\end{align}
implying
\begin{align}
    \frac{1}{n}F\Phi &= X.
\end{align}
We may similarly recover $\frac{1}{n}F \y = \hat{\y}$ to show that the coefficients $\hat{y}_k$ are given by a discrete Fourier transform of $y_j$.

\subsubsection{Error analysis of the linear model}

Having shown that the discrete Fourier transform $F$ relates the original system of trigonometric equations $\Phi \balpha = \y$ to the system of linear equations (or ordinary least squares  problem $X \balpha = \hat{\y}$, where we treat the rows of $X$ as observations in $\R^d$), we now derive the error of the problem in the Fourier representation. Standard treatment for ordinary least squares (OLS) gives the minimum $\ell_2$-norm solution to Eq.~\eqref{eq:alpha_opt_prob_fourier} as
\begin{equation}\label{eq:alpha_opt_lin}
    \balpha^{opt} = X^T (X X^T)^{-1} \hat{\y}
\end{equation}
in which case the optimal interpolating overparameterized Fourier model is
\begin{align}\label{eq:class_solved}
    f^{opt}(x) &= \langle \phi(x), \balpha^{opt}\rangle
    \\&=  \sum_{k \in \Omega_n} \frac{\hat{y}_k}{\sum_{j \in S(k)} \nu_j } \, \sum_{\ell\in S(k)}  \nu_{\ell} e^{i 2\pi \ell x}.
\end{align}
Once we have trained a model on noisy samples $\y$ of $g$ using uniformly spaced values of $x$, we would like to evaluate how well the model performs for arbitrary $x$ in $[0, 1]$. Given some function $f = \langle \phi(x), \balpha\rangle$ the mean squared error $\E_x(f(x) - g(x))^2$ may be evaluated with respect to the interval $[0, 1]$. We define the \textit{generalization error} of the model as the expected mean squared error evaluated with respect to $\y$:
\begin{align}
    L(f) := \E_{x, \y} (f(x) - g(x))^2.
\end{align}
We decompose the generalization error of the model as
\begin{align}
    \E_{x, \mathbf{y}} (f(x) - g (x) )^2 &= \E_{x, \mathbf{y}} (f(x) - \E_\mathbf{y} f(x) + \E_\mathbf{y} f(x) - g (x) )^2
    \\&= \E_{x, \mathbf{y}} (f(x) - \E_\mathbf{y} f(x))^2 + \E_{x, \mathbf{y}} ( \E_\mathbf{y} f(x) - g (x) )^2 
    \\ &\qquad \qquad + 2 \E_{x, \mathbf{y}} \left[ (f(x) - \E_\mathbf{y} f(x))( \E_\mathbf{y} f(x) - g (x) )\right]. \nonumber 
\end{align}
Because of orthonormality of Fourier basis functions, the cross-terms cancel resulting in a decomposition of the generalization error of the optimal standard bias and variance terms:
\begin{align}
   L(f^{opt}) &= \underbrace{\E_{x, \mathbf{y}} (f^{opt}(x) - \E_\mathbf{y} f^{opt}(x))^2}_{\textsc{var}} + \underbrace{\E_{x, \mathbf{y}} ( \E_\mathbf{y} f^{opt}(x) - g (x) )^2}_{\textsc{bias}^2}.
\end{align}
We now evaluate $\textsc{var}$ and $\textsc{bias}^2$ using the linear representation developed in the previous section. Beginning with the variance, conditional on constructing $\Phi$ from the set of uniformly spaced points $x_j$ we apply the discrete Fourier transformation to yield $X$ and compute $\balpha^{opt}$ using Eq.~\eqref{eq:alpha_opt_lin}:
\begin{align}
    \textsc{var} &= \E_{x, \mathbf{y}} (f^{opt}(x) - \E_\mathbf{y} f^{opt}(x))^2
    \\&=\E_{x, \mathbf{y}} |\langle \phi(x), \balpha^{opt}- \E_\mathbf{y} \balpha^{opt} \rangle |^2
    \\&= \E_{x, \mathbf{y}} |\langle \phi(x), X^T (X X^T)^{-1} (\hat{\y} - \E_\mathbf{y} \hat{\y}) \rangle |^2
    \\&=  \tr{ \E_{\y}\left[(\hat{\y} - \E_\mathbf{y} \hat{\y})(\hat{\y} - \E_\mathbf{y} \hat{\y})^\dagger \right]  (X X^T)^{-1} X \E_x\left[\phi(x) \phi(x)^\dagger \right] X^T   (X X^T)^{-1}  }
    \\&= \frac{\sigma^2}{n}  \tr{  (X X^T)^{-2} X \Sigma_\phi X^T  }.\label{eq:var_1n}
\end{align}
Letting $\boldsymbol{\epsilon} := (\y - \E_\y \y)$ we have simplified the above using the following\footnote{In Eq.~\eqref{eq:var_1n}, $\E[X^T X] = \Sigma_\phi$, so that $\Sigma_\phi$ differs from the covariance matrix for the rows of $X$ by a factor of $1/n$.}:
\begin{align}
    \E_{\y}\left[(\hat{\y} - \E_\mathbf{y} \hat{\y})(\hat{\y} - \E_\mathbf{y} \hat{\y})^\dagger \right]_{jk} &= \frac{1}{n^2}   \E_{\y} (F \boldsymbol{\epsilon})_j(F \boldsymbol{\epsilon})_k^*
    \\&=   \frac{1}{n^2}\E_{\y} \left( \sum_{p\in\Omega_n} \omega^{-jp} \epsilon_p \right)\left( \sum_{q\in\Omega_n} \omega^{qk} \epsilon_q \right)
    \\&= \frac{1}{n^2}\sum_{p,q\in\Omega_n} \E_\y [\epsilon_p \epsilon_q]  \omega^{-jp+kq}
    \\&= \frac{\sigma^2}{n} \delta_j^k,
\end{align}
where we have used the fact that the errors $\epsilon$ are independent and zero mean, $\E_\y [\epsilon_p \epsilon_q] = \E_\y[\epsilon_p^2] \delta_p^q$. We have defined the feature covariance matrix as $\Sigma_\phi = \E_x [\phi(x) \phi(x)^\dagger]$, which may be computed elementwise using the orthonormality of Fourier features on $[0, 1]$:
\begin{align}
    (\Sigma_\phi)_{jk} = \E_{x} [\phi(x)_j \phi(x)_k^*] = \sqrt{\nu_j \nu_k} \langle e^{i 2\pi j x}, e^{i 2\pi kx} \rangle_{L_2^{[0,1]}} = \delta_j^k \nu_j.
\end{align}
The following may be computed directly:
\begin{align}
    (XX^T)_{jk} &= \delta_j^k \sum_{\ell \in S(j) } \nu_\ell\\
    (X \Sigma_\bx X^T)_{jk} &= \sum_{\ell\in \Omega_d} X_{j\ell} \left(\frac{\nu_\ell}{n}\right)X_{k\ell} 
    \\&= \frac{1}{n}\sum_{\ell\in \Omega_d}  \nu_p^2  \cond{\ell \in S(j)}  \cond{\ell \in S(k)}
    \\&=  \frac{\delta_j^k }{n}\sum_{\ell \in S(k)} \nu_\ell^2. \label{line:ident}
\end{align}
In line~\eqref{line:ident} we have used the identity
\begin{equation}\label{eq:double_cond}
    \cond{\ell \in S(j)}  \cond{\ell \in S(k)} =  \cond{\ell \in S(j)} \delta_k^j,
\end{equation}
since $\ell \in S(k) \Rightarrow k = \ell \text{ mod } n$ and therefore $\ell \in S(j) \Rightarrow j =  \ell \text{ mod } n = k$.  We now compute the variance as
\begin{align}
    \textsc{var} &= \frac{\sigma^2}{n} \tr{(XX^T)^{-2} X \Sigma_\phi X^T)} \\
    &=  \frac{\sigma^2}{n}\sum_{j,k \in \Omega_n} \left(\delta_k^j \sum_{\ell \in S(k) } \nu_\ell \right)^{-2} \left(\delta_j^k \sum_{\ell \in S(k)} \nu_\ell^2 \right)
    \\&=  \frac{\sigma^2}{n}\sum_{k \in \Omega_n} \frac{\sum_{\ell \in S(k)} \nu_\ell^2}{\left(\sum_{\ell \in S(k) } \nu_\ell \right)^{2}}. \label{eq:var_v2} 
\end{align}
To evaluate the $\textsc{bias}^2$, we will first rewrite $g(x)$ in terms of its Fourier representation,
\begin{align}
    g(x) &= \langle \phi(x), \balpha_0\rangle, \\
    (\balpha_0)_k &= \frac{\hat{g}_k}{\sqrt{\nu_k}} \cond{k \in \Omega_{n_0}}.
\end{align}
Noting that $(X\balpha_0)_k = \hat{g}_k$ implies $X \balpha_0 = \E_\y \hat{\y}$ we evaluate
\begin{align}
    \textsc{bias}^2 &= \E_{x, \y} ( \E_\mathbf{y} f^{opt}(x) - g (x) )^2 
    \\&=\E_{x, \y} |\langle \phi(x), \E_\y \balpha^{opt} - \balpha_0\rangle|^2
    \\&= \tr{\E_x \left[\phi(x)\phi(x)^\dagger \right](\E_\y \balpha^{opt} - \balpha_0)(\E_\y \balpha^{opt} - \balpha_0)^\dagger}
    \\&= \tr{ \Sigma_\phi(X^T (X X^T)^{-1} X  - \I_d ) \balpha_0\balpha_0^\dagger(X^T (X X^T)^{-1} X  - \I_d )^\dagger }
    \\&= \norm{\Sigma_\phi^{1/2}  (\I_d  -X^T (X X^T)^{-1} X ) \balpha_0}_2^2. \label{eq:bias_v2_pre}
\end{align}
While Eq.~\eqref{eq:bias_v2_pre} already completely characterizes the bias error in terms of the choice of \fweight{s} and input data, it may be greatly simplified by taking advantage of the sparseness $X$. We have
\begin{align}
    (X^T (X X^T)^{-1} X)_{jk} &= \sum_{\ell \in \Omega_n} \left(\sum_{m \in S(\ell) } \nu_m \right)^{-1} X_{\ell j} X_{\ell k}
    \\&=  \left(\sum_{m \in S(j \text{ mod } n) } \nu_m \right)^{-1} \sqrt{\nu_k} \sqrt{\nu_j} \cond{k \in S(j \text{ mod } n)}, \label{line:iden2}
\end{align}
where in line~\eqref{line:iden2} we have used
\begin{equation}
   \sum_{\ell \in \Omega_n} \cond{k \in S(\ell)} \cond{j \in S(\ell)} = \cond{k \in S(j \text{ mod } n)}
\end{equation}
for $k,j \in \Omega_d$ and $\ell \in \Omega_n$, which follows from similar reasoning as Eq.~\eqref{line:ident}. And so, writing 
\begin{align}
     (\Sigma_\phi^{1/2}& (\I_d - X^T (X X^T)^{-1} X) \balpha_0)_\ell 
     \\&=   \sum_{j \in \Omega_d} \Bigg[\delta_\ell^j \hat{g}_j \cond{j \in \Omega_{n_0}}    \nonumber
     \\&\qquad  - \sum_{k \in \Omega_d} \sqrt{\nu_\ell} \delta_j^{\ell} \left(\sum_{m \in S(j \text{ mod } n) } \nu_m \right)^{-1} \hat{g}_k \sqrt{\nu_j} \cond{k \in S(j \text{ mod } n)}  \cond{k \in \Omega_{n_0}}  \Bigg] 
     \\&=  \left( \sum_{j \in \Omega_{n_0}} \delta_\ell^j \hat{g}_j -  \sum_{k \in \Omega_{n_0}} \nu_\ell  \left(\sum_{m \in S(\ell\text{ mod } n) } \nu_m \right)^{-1} \hat{g}_\ell \cond{k \in S(\ell \text{ mod } n)} \right)
     \\&=  \left( \cond{\ell \in \Omega_{n_0}} \hat{g}_j -  \sum_{k \in \Omega_{n_0}} \nu_\ell  \left(\sum_{m \in S(k) } \nu_m \right)^{-1} \hat{g}_k \cond{\ell \in S(k)}  \right).
\end{align}
Letting $Q_k := \sum_{m \in S(k) } \nu_m $ the bias term of the error evaluates to
\begin{align}
     \textsc{bias}^2 &= \sum_{\ell \in \Omega_d} |(\Sigma_\bx^{1/2} (\I_d - X^T (X X^T)^{-1} X) \balpha_0)_\ell|^2 
     \\&=   \sum_{\ell \in \Omega_d} \left| \cond{\ell \in \Omega_{n_0}} \hat{g}_\ell -  \sum_{k \in \Omega_{n_0}} \nu_\ell Q_k^{-1}  \hat{g}_k \cond{\ell \in S(k)} \right|^2
     \\&=  \sum_{\ell \in \Omega_d} \left(  \cond{\ell \in \Omega_{n_0}} |\hat{g}_\ell|^2 -  \sum_{k \in \Omega_{n_0}}\cond{\ell \in \Omega_{n_0}} \hat{g}_\ell \hat{g}_k^* \nu_\ell Q_k^{-1}  \cond{\ell \in S(k)} \right. \nonumber
     \\&\qquad\qquad\left. - \sum_{k\in\Omega_{n_0}}\cond{\ell \in \Omega_{n_0}} \hat{g}_\ell^* \hat{g}_k  \nu_\ell Q_k^{-1}  \cond{\ell \in S(k)}  \right. \nonumber 
     \\ &\qquad\qquad\left.+ \sum_{j,k\in\Omega_{n_0}} \nu_\ell^2 Q_j^{-1} Q_k^{-1} \hat{g}_j^* \hat{g}_k \cond{\ell \in S(j)} \cond{\ell \in S(k)}  \right)
     \\&=  \left( \sum_{k \in \Omega_n} |\hat{g}_k|^2 - 2\sum_{k \in \Omega_{n_0}}  \nu_k  |\hat{g}_k|^2 Q_k^{-1} + \sum_{k \in \Omega_{n_0}} \sum_{\ell \in S(k)} \nu_\ell^2 Q_k^{-2} |\hat{g}_k|^2 \right)
    \\&= \sum_{k\in\Omega_{n_0}} \frac{|\hat{g}_k|^2}{Q_k^2} \left( (Q_k - \nu_k )^2 + \sum_{\ell \in S(k)\backslash k} \nu_\ell^2 \right)
    \\&=  \sum_{k\in\Omega_{n_0}}  |\hat{g}_k|^2 \frac{\left(\sum_{\ell \in S(k)\backslash k } \nu_\ell \right)^{2}  + \sum_{\ell \in S(k)\backslash k} \nu_\ell^2}{\left(\sum_{\ell \in S(k) } \nu_\ell \right)^{2}}. \label{eq:bias_v2}
\end{align}

With Eqs.~\eqref{eq:var_v2} and \eqref{eq:bias_v2} we have recovered a closed form expression for generalization error in terms of \fweight{s} $\nu_k$, the target function Fourier coefficients $\hat{g}_k$, and the noise variance $\sigma^2$. This was possible in part because of the orthonormality of the Fourier features $\phi(x)_k$ and the choice to sample $x_j$ uniformly on $[0, 1]$, resulting in diagonal $\Sigma_\phi$ and $X X^T$ respectively. This simplicity is more advantageous for studying scenarios where benign overfitting may exist compared to prior works \cite{bartlett2020benign,tsigler2020}. We will now analyze choices of \fweight{s} $\nu_k$ that may give rise to benign overfitting for overparameterized weighted Fourier Features models.

We remark that the results of this section may also be derived using the methods of Ref.~\cite{muthukumar2020harmless}, though we have opted here to use a language more reminiscent of linear regression to highlight connections between the analysis of weighted Fourier features models and ordinary least squares errors.

\subsection{Solutions for special cases}\label{app:special_cases}

\subsubsection{Noise-only minimum norm estimator}\label{sec:zerosignal}

We now derive the cases considered in Sec.~\ref{sec:cases} of the main text. We first consider when the target function is given by $g = 0$ and we attempt to predict on a pure noise signal
\begin{equation}
    y_j = \epsilon_j,
\end{equation}
with $\E[\epsilon^2]= \sigma^2$. We can recover an unaliased, ``simple'' fit to this pure noise signal by reconstructing $\boldsymbol{\epsilon}$ from the DFT using Eq.~\eqref{eq:dft}:
\begin{equation}
    \epsilon_j = \sum_{k \in \Omega_n} \hat{\epsilon}_k e^{i 2\pi jk/n}.
\end{equation}
By setting all weights $\nu_k = 1$, an immediate choice for an interpolating $f$ with access to frequencies in $\Omega_d$ is found by setting 
\begin{equation}\label{eq:case1_sol}
    \alpha_k = \frac{\hat{\epsilon}_{k\text{ mod } n}}{n(m+1)},
\end{equation}
where we have assumed $d = n (m+1)$. Eq.~\eqref{eq:case1_sol} is the minimum $\ell_2$-norm estimator and evenly distributes the weight of each Fourier coefficient $\hat{\epsilon}_k$ over $m$ aliases frequencies in in $S(k)$. The effect of higher-frequency aliases is to reduce the $f$ to near-zero everywhere except at the interpolated training data. We can directly compute the generalization error of the interpolating estimator as 
\begin{align}
     \E_{x,\y} |f(x) - g(x)|^2 &= \E_x \left|f(x) \right|^2 
     \\ &= \E_\y \left| \sum_{k=0}^{d-1} \alpha_k e^{-i 2\pi k x} \right|^2 \label{eq:line00}
    \\&= \E_\y \sum_{k=0}^{d-1} \left|\frac{\hat{\epsilon}_{k\text{ mod } n}}{n(m+1)}\right|^2\label{eq:line01}
    \\&= \E_\y \frac{n(m+1)\norm{\hat{\boldsymbol{\epsilon}}}_2^2}{n^2(m+1)^2}\label{eq:line02}
    \\&=  \frac{\sigma^2}{n(m+1)}. \label{eq:line03}
\end{align}
Using $d = n(m+1)$ as the dimensionality of the feature space, we have recovered the lower bound scaling for overparameterized models derived in Ref.~\cite{muthukumar2020harmless}. In line~\eqref{eq:line00} we have used the independence of $\epsilon$ from $x_k$ and $y_k$ and in line~\eqref{eq:line01} we have used orthonormality of Fourier basis functions on $[0, 1]$. line~\eqref{eq:line03} uses Parseval's relation for the Fourier coefficients, namely:
\begin{equation}
 \sum_{k \in \Omega_n} |\epsilon_k|^2  = n \sum_{k \in \Omega_n} |\hat{\epsilon}_k|^2,
\end{equation}
which implies $
\E \norm{\hat{\boldsymbol{\epsilon}}}_2^2 = n^{-1}\E\norm{\boldsymbol{\epsilon}}_2^2 = \sigma^2$. Figure~\ref{fig:noise_noiseless_demo}a of the main text shows the effect of the number of cohorts $m$ on the behavior of $f$, which interpolates a pure noise signal with very little bias. As the number of cohorts increases, the function deviates very little away from the true ``signal'' $y=0$, and becomes very ``spiky'' in the vicinity of noise

\subsubsection{Signal-only minimum norm estimator}\label{sec:zeronoise}

Now we study the opposite situation in which the pure tone is noiseless, and aliases in the spectrum of $f$ interfere in prediction of $f$. In this case, we set $\sigma = 0$ and interpolate target labels
\begin{align}
     y_j &= \hat{g}_p e^{i 2\pi p x_j},
\end{align}
with $-n/2 < p < n/2$. When we set $d = n(m+1)$ and predict on $y$, there are exactly $|S(p)|-1 = m$ aliases for the target function with frequency $p$. We again assume all \fweight{s} are equal, $\nu_k = 1$, and by orthonormality of Fourier basis functions, only the components of $f$ with frequency in $S(p)$ are retained:
\begin{equation}
    f(x) = \alpha_{p} e^{i2\pi p x} + \sum_{k\in S(p)} \alpha_k e^{i2\pi kx},
\end{equation}
which will interpolate the training points for any choice of $\boldsymbol{\alpha}$ satisfying 
\begin{equation}\label{eq:interp1}
    \sum_{k \in S(p)} \alpha_k = \hat{g}_p.
\end{equation}
The choice of \tcoef{s} $\alpha_k$ that satisfy Eq.~\eqref{eq:interp1} while minimizing $\ell_2$-norm is
\begin{equation}
    \alpha_k = \begin{cases}
    \frac{\hat{g}_p}{m+1}, & k \in S(p) ,\\
    0, & \text{ otherwise}.
    \end{cases}
\end{equation}
The problem with minimizing the $\ell_2$ norm in this case is that it spreads the true signal into higher frequency aliases: The generalization error of this model is 
\begin{align}
    \E_{x, \y} |f(x) - g(x)|^2 = \left| \hat{g}_p \left( \frac{m}{m+1}\right)\right|^2 +  m\frac{|\hat{g}_p|^2}{(m+1)^2} = \mathcal{O}(1).
\end{align}
 We see that this model generally fails to generalize. This poor generalization was described as ``signal bleeding'' by Ref.~\cite{muthukumar2020harmless}: Using $n$ samples, there is no way to distinguish the signal due to an alias of $p$ from the signal due to the true frequency $p$, so the coefficients $\alpha_k$ become evenly distributed over aliases with very little weight allocated to the true Fourier coefficient $\hat{g}_{p}$ in the model $f$. Fig.~\ref{fig:noise_noiseless_demo}b in the main text shows the effect of ``signal bleed'' for learning a pure tone in the absence of noise.

\subsection{Conditions for Benign overfitting}\label{app:classical_bo}

The behavior of the error of the overparameterized Fourier model (Eq.~\eqref{eq:class_err} of the main text) depends on an interplay between noise variance $\sigma^2$, signal Fourier coefficients $\hat{g}_k$, \fweight{s} $\nu_k$, and the size of the model $d$. A desirable property of the \fweight{s} $\nu_k$ is that they should result in a model that both interpolates the sampled data while also achieving good generalization error in the limit $n\rightarrow \infty$. For our purposes we will consider cases where $\lim_{n\rightarrow \infty } L(f^{opt}) = 0$, though this condition could be relaxed to allow for more interesting or natural choices of $\nu_k$. We now analyze the error arising from a simple weighting scheme to demonstrate benign overfitting using the overparameterized Fourier models discussed in this work.

\subsubsection{A simple demonstration of benign overfitting}\label{sec:simple_bo}

Here we demonstrate a simple example of benign overfitting when the \fweight{s} $\nu_k$ are chosen with direct knowledge of the spectrum $\Omega_{n_0}$ of $g$. For some $n_0 < n < d$, fix $c \in (0, 1)$ and use the \fweight{s} given as
\begin{align}\label{eq:hatweights}
    \nu_k = \begin{cases}
      \frac{c}{n_0}, & k \in \Omega_{n_0}, \\
      \frac{1-c}{d-n_0}, & \text{otherwise},
    \end{cases}
\end{align}
for all $k \in \Omega_d$. For simplicity, suppose $d = n(m+1)$ such that $|S(k)| = m + 1$ for all $k \in \Omega_n$. Defining the \textit{signal power} as 
\begin{equation}\label{eq:signalpower}
    P := \sum_{k \in \Omega_{n_0}} |\hat{g}_k|^2,
\end{equation}
we can directly evaluate \textsc{var} of Eq.~\eqref{eq:var_v2} and $\textsc{bias}^2$ of Eq.~\eqref{eq:bias_v2}:
\begin{align}
    \textsc{var} &\rightarrow \frac{\sigma^2}{n}\left( n_0  \frac{\left(\frac{c}{n_0}\right)^2 + m \left( \frac{1-c}{d - n_0} \right)^2}{\left(\frac{c}{n_0} + m\frac{1-c}{d - n_0}  \right)^2} + \frac{n-n_0}{m+1}\right) . \\
    \textsc{bias}^2 &\rightarrow  \frac{P n_0^2  (m+1)m }{\left(\frac{c}{1-c}(d - n_0) + m n_0 \right)^2} .
\end{align}
Fixing $n_0$, we can bound generalization error in the asymptotic limit $n \rightarrow \infty$ as:
\begin{align}
    \textsc{var} &= \mathcal{O}\left( \frac{1}{n} + \frac{n}{d}\right)\\
    \textsc{bias}^2 &= \mathcal{O} \left( \frac{1}{n^2}\right).
\end{align}
Therefore, by setting $d = \omega(n)$ the model perfectly interpolates the training data and also achieves vanishing generalization error in the limit of large number of data. A similar example was considered in Ref.~\cite{muthukumar2020harmless}, though a rigorous error analysis (and relationship to benign overfitting) was not considered there. This benign overfitting behavior is entirely due to the \fweight{s} of Eq.~\eqref{eq:hatweights}: As $d,n \rightarrow \infty$ with $d = \omega(n)$, the \fweight{s} are concentrated on $\Omega_{n_0}$ (suppressing $\textsc{bias}$) while becoming increasingly small and evenly distributed over all aliases of $\Omega_{n_0}$ (suppressing $\textsc{var}$).

\subsubsection{More general conditions for benign overfitting}\label{sec:general_bo}

We have derived closed-form solutions for the $\textsc{bias}^2$ and $\textsc{var}$ terms that determine the total generalization error of an interpolating model $f$ and in the previous section provided a concrete example of a model that achieves benign overfitting in this setting. We now discuss conditions under which a more general choice of model can exhibit benign overfitting.

We begin by showing that the variance of Eq.~\eqref{eq:var_v2} splits naturally into an error due to a (simple) prediction component and a (spiky, complex) interpolating component. Following Refs.~\cite{bartlett2021deep,bartlett2020benign}, we will split the variance into components corresponding to eigenspaces of $\Sigma_\phi$ with large and small eigenvalues respectively. Let $S^{\leq p}$ denote the set indices for the largest $p$ eigenvalues of $\Sigma_\phi$ (i.e., the largest $p$ values of $\nu_k$), and $S^{>p} = [d] \backslash S^{\leq p}$ be its complement. Define $P^{\leq p}: \R^d \rightarrow \R^d$ as the projector onto the the subspace of $\R^d$ spanned by basis vectors labelled by indices in $S^{\leq p}$ (and $P^{>p} = \I_d - P^{\leq p}$ is defined analogously). Then letting $(\bnu)_k = \nu_k$ be the vector of \fweight{s} and assuming $p \leq n$, we may rewrite the variance of Eq.~\eqref{eq:var_v2} as
\begin{align}
    \textsc{var} &= \frac{\sigma^2}{n}\sum_{k \in \Omega_n} \frac{\sum_{\ell \in S(k)} \nu_\ell^2}{\left(\sum_{\ell \in S(k) } \nu_\ell \right)^{2}}
    \\&= \frac{\sigma^2}{n}\sum_{k \in \Omega_n} \frac{\norm{P_{S(k)} \bnu}_2^2}{\norm{P_{S(k)} \bnu}_1^2}
    \\&=  \frac{\sigma^2}{n}\sum_{k \in \Omega_n} \frac{\norm{P^{\leq p}P_{S(k)}\bnu + P^{>p} P_{S(k)} \bnu}_2^2}{\norm{P_{S(k)} \bnu}_1^2}
    \\&\leq  \frac{\sigma^2}{n}\left( \sum_{k \in \Omega_n} \frac{\norm{P^{\leq p}P_{S(k)}\bnu}_2^2}{\norm{P^{\leq p} P_{S(k)}\bnu}_1^2} +\sum_{k \in \Omega_n} \frac{\norm{P^{>p} P_{S(k)} \bnu}_2^2}{\norm{P^{>p} P_{S(k)} \bnu}_1^2} \right)
    \\&\leq \sigma^2 \left[\frac{p}{n} + \frac{1}{n}\sum_{k \in \Omega_n} \frac{1}{R_p^{(k)}(\Sigma_\phi)} \right],
\end{align}
where we have used $P^{\leq p}P_{S(k)}\hat{e}_j = \delta_j^k \cond{j \leq p}$ for $p \leq n$ and we have introduced an effective rank for the alias cohort of $k$,
\begin{align}
R_p^{(k)}(\Sigma_\phi) = \frac{\left(\sum_{\ell \in S(k) \cap S^{>p} } \nu_\ell \right)^2}{\sum_{\ell \in S(k) \cap S^{>p}} \nu_\ell^2}.
\end{align}
Since $p=n_0$ is a relevant choice for our problem setup, we define $R^{(k)} := R_{n_0}^{(k)}(\Sigma_\phi)$ and focus on the bound 
\begin{align}
    \textsc{var} &\leq \sigma^2 \left[\frac{n_0}{n} + \frac{1}{n}\sum_{k \in \Omega_n} \frac{1}{R^{(k)}} \right],\label{eq:var_ub}
\end{align}
The first term of the decomposition of Eq.~\eqref{eq:var_ub} corresponds to the variance of a (non-interpolating) model with access to $n_0$ Fourier modes, while the second term corresponds to excess error introduced by high frequency components of $f$. Given a sequence of experiments with increasing $n$ (while $g$ and $\sigma^2$ remain fixed), we would like to understand the choices of \fweight{s} $\nu_k$ for which $\textsc{var}$ vanishes as $n\rightarrow \infty$. Given that $|\{k\in \Omega_n\}| = n$, a sufficient condition for a sequence of \fweight{} distributions to be benign is that $R^{(k)} = \omega(1)$ for all $k$ while a necessary condition is that there is no $k$ for which $R^{(k)} = \mathcal{O}(1/n)$. These conditions are not difficult to satisfy: Intuitively they require only that $\Sigma_\phi$ changes with increasing $n$ in such a way that the values of $\nu_k$ in $\Omega_n \backslash \Omega_{n_0}$ continue ``flatten out'' as $n$ increases. This is precisely the behavior engineered in the example of Sec.~\ref{sec:simple_bo}.

We now proceed to bound the bias term of Eq.~\eqref{eq:bias_v2_pre}. Observe that $P^\perp := (\I_d - X^T (X X^T)^{-1} X)$ is a projector onto the subspace of $\R^d$ orthogonal to the rows of $X$, therefore satisfying $\norm{P^\perp} \leq 1$ and $P^\perp X^T X = 0$. Then the $\textsc{bias}^2$ is bounded as
\begin{align}
    \textsc{bias}^2 &= \norm{\Sigma_\phi^{1/2} P^\perp \balpha_0}_2^2
    \\&= \norm{(\Sigma_\phi - X^T X)^{1/2} P^\perp \balpha_0}_2^2
    \\&\leq \norm{\Sigma_\phi - X^T X}_\infty \norm{\balpha_0}_2^2. \label{eq:bias_ub}
\end{align}
The term $X^T X $ can be interpreted as a finite-sample estimator for $\Sigma_\phi$ in the sense that $X^T X  = n^{-1} \Phi^T \Phi = \Sigma_\phi$ for $n=d$. However, we cannot apply standard results on the convergence of sample covariance matrices (e.g. Ref.~\cite{Koltchinskii_2010}) since the uniform spacing requirement for training data $x$ violates the standard assumption of i.i.d. input data. To proceed, we will make a number of simplifying assumptions about the \fweight{s}. First, to control for the possibility that large \fweight{s}  $\nu_k$ concentrate within a specific $S(k)$ we will assume that \fweight{s} corresponding to any set of alias frequencies of $\Omega_n$ are close to their average. Letting $d = (m+1)n$, we  define 
\begin{align}
    \eta_\ell = \frac{1}{n} \sum_{k \in \Omega_n} \nu_{k + n\ell},
\end{align}
for $\ell =0, 1, \dots, m$. We will impose that $| \nu_{k + n\ell} - \eta_\ell| \leq t$ for all $k \in \Omega_n$, $\ell =1, \dots, m$, for some positive number $t$. We further assume normalization, $\sum_{k \in \Omega_d} \nu_k = 1 = n\sum_{\ell=0}^m \eta_\ell$. Under these assumptions we can bound the first term of \eqref{eq:bias_ub} as
\begin{align}
    \norm{\Sigma_\phi - X^T X}_\infty &\leq \max_j \nu_j^{1/2} \left( \sum_{k \in S(j\text{ mod } n) \backslash j} \nu_k^{1/2} \right) \label{line:gershgorin}
    \\&\leq (m-1)^{1/2}\max_j \nu_j^{1/2} \left( \sum_{k \in S(j\text{ mod } n) \backslash j} \nu_k \right)^{1/2}
    \\&\leq  m^{1/2}\max_j \nu_j^{1/2} \left( \sum_{\ell=1}^m (\eta_\ell + t)\right)^{1/2}
    \\&\leq \left(\frac{m}{n} + m^2 t\right)^{1/2}, \label{eq:spec_ub}
\end{align}
where in line~\eqref{line:gershgorin} we have used Gershgorin circles. Meanwhile, defining $\zeta := \left( \sum_{k \in \Omega_{n_0}} \nu_k \right)$, by Cauchy-Schwarz we have that
\begin{equation}\label{eq:alpha_ub}
    \norm{\balpha_0}^2 \leq P  \zeta^{-2},
\end{equation}
where $P$ is the signal power of Eq.~\eqref{eq:signalpower}. A necessary condition for producing benign overfitting in overparameterized Fourier models is that $\zeta$ remains relatively large as $n,d\rightarrow \infty$. If this is accomplished, then a small enough $t$ guarantees that all \fweight{s} associated with frequencies $k \in \Omega_d \backslash \Omega_{n_0}$ will be uniformly suppressed. For instance, if $\zeta$ is lower bounded as a constant while $t=0$ then combining Eqs.~\eqref{eq:spec_ub} and~\eqref{eq:alpha_ub} yields a bound of
\begin{equation}
    \textsc{bias}^2 = \mathcal{O}\left(\frac{d^{1/2}}{n}\right).
\end{equation}
Although the analysis of Sec.~\ref{sec:simple_bo} yields a significantly tighter bound, this demonstrates that the mechanisms behind that simple demonstration of benign overfitting are somewhat generic. In particular, normalization and lower bounded support of the \fweight{s} on $\Omega_{n_0}$ is almost sufficient to control the bias term of the generalization error.

\section{Solution for the quantum overparameterized model}\label{app:quantum_opt_err}

We now derive the solution to the minimum-norm interpolating quantum model, 
\begin{align}
    M^{opt} &= \argmin_{\substack{M = M^\dagger \\ f(x_j) = y_j}} \norm{M}_F, \\
    f(x) &= \sum_{k \in \Omega} e^{i 2\pi k x} \sum_{\ell, m \in R(k)} \gamma_\ell M_{m \ell} \gamma_m^*. \label{eq:qmodel_2}
\end{align}
We will use the following notation and definitions:
\begin{align}
    a_k &= -\Biggl\lfloor  \frac{ k + \frac{|\Omega| - n}{2} }{n} \Biggl\rfloor, \\
    b_k &= \Biggr\lfloor  \frac{ -k + \frac{|\Omega| + n}{2} - 1 }{n} \Biggr\rfloor ,\\
    S_k &= \{k - a(k) n, \dots, k, k+n, \dots, k+b(k)n\} \\&\qquad \qquad= \{k+pn, \quad p=-a(k), \dots, b(k) \}, \\
    R(S_k) &:= \bigcup_{j \in S_k} R(j). 
\end{align}
Here, $a_k$ and $b_k$ characterize the number positive and negative frequency aliases of $k$ appearing in $\Omega$ (i.e., $k + n\ell \in \Omega$ for all $a_k \leq \ell \leq b_k$) assuming that $|\Omega|$ is odd, a requirement for any quantum model. Let $L(d)$ be the space of linear operators acting on $d \times d$ matrices. Define the linear operator $P_k: L(d) \rightarrow L(d)$ for $k \in [n]$ as
\begin{equation}
    P_k(X) = \sum_{\ell,m \in R(k)} \langle \ell | X | m \rangle \kb{\ell}{m}
\end{equation}
for any $X \in L(d)$. Importantly, $P_k$ is not necessarily Hermitian preserving. Denoting $\Gamma := \dm{\Gamma}$ for brevity, we may rewrite the Fourier coefficients of $f(x)$ as
\begin{align}
    \langle P_k(M), P_k(\Gamma) \rangle &= \tr{P_k(M)^\dagger P_k(\Gamma)} 
    \\&= \tr{\sum_{\ell,m \in R(k)}M_{\ell m}^* \kb{m}{\ell} \sum_{i,j \in R(k)} \gamma_i \gamma_j^* \kb{i}{j}}
    \\&= \sum_{\ell,m\in R(k)} \gamma_\ell^* M_{m\ell} \gamma_m,
\end{align}
where in the last line we have used hermiticity of $M$. Applying $f(x_j) = y_j$ $\forall \, j \in [n]$ and substituting into Eq.~\eqref{eq:qmodel_2} we find the interpolation condition
\begin{align}
    \hat{y}_p &= \sum_{q=a_k}^{b_k} \langle P_{p + nq}(M), P_{p + nq}(\Gamma) \rangle 
    \\&= \langle P_{S_p}(M), P_{S_p}(\Gamma)\rangle , \qquad \forall \, p \in [n],
\end{align}
where $P_{S_p} := \sum_{q=a_k}^{b_k} P_{p + nq}$. The equality follows from
\begin{equation}
    \langle P_j(X), P_k(Y)\rangle = \langle P_j(X), P_j(Y)\rangle \delta_j^k
\end{equation}
due to the fact that $R(j)$ and $R(k)$ are disjoint sets for any $j \neq k$. Following the technique of Ref.~\cite{muthukumar2020harmless} we apply the Cauchy-Schwarz inequality to find
\begin{equation}\label{eq:CS}
    \norm{P_{S_p}(M)}_F \geq  \frac{|\langle P_{S_p}(M), P_{S_p}(\Gamma) \rangle|}{\norm{P_{S_p}(\Gamma)}_F},
\end{equation}
with equality if and only if $P_{S_p}(M)$ is proportional to $P_{S_p}(\Gamma)$. Saturating this lower bound by setting $M_{\ell m} = c \gamma_\ell \gamma_m^*$ and solving for the proportionality $c$ constant using Eq.~\eqref{eq:CS}, we find
\begin{align}
    \hat{y}_p = \langle c P_{S_p}(\Gamma), P_{S_p}(\Gamma)\rangle  =  \sum_{\ell, m \in R(S_p)} c^*|\gamma_\ell|^2 |\gamma_m|^2. 
\end{align}
This indicates an additional requirement for interpolation that $\gamma_\ell,\gamma_m > 0$ for some pair $(\ell, m)\in R(k)$ whenever $\tilde{y}_k \neq 0$ and so for simplicity we will require that $\gamma_\ell > 0$ for all $\ell = 1, \dots, d$. Within each set of indices $R(S_k)$, the elements of the optimal observable are defined piecewise with respect to that partition $R$:
\begin{equation}\label{eq:Mopt_piecewise}
    (\ell, m) \in R(S_k) \Rightarrow M_{\ell m} = \hat{y}_k^* \frac{\gamma_\ell \gamma_m^*}{\sum_{i,j \in R(S_k)} |\gamma_i|^2 |\gamma_j|^2} .
\end{equation}
Minimization of $\norm{M}_F$ subject to the interpolation constraint is equivalent to minimization of $\norm{P_{p+nq}(M)}_F$ for all $q\in [a_k, b_k], \,k \in \Omega$, and so solving the constrained optimization over all distinct subspaces in $\bigcup_{k=0}^{n-1} R(S_k) = \{0,1\}^n \times \{0,1\}^n $ we recover the optimal observable
\begin{equation}\label{eq:Mopt_hard}
    M_{opt} = \argmin_{f(x_j)=y_j \,\forall\, j} \norm{M}_F = \sum_{k \in \Omega_n} \hat{y}_k^* \sum_{\ell, m \in R(S_k)} \frac{\gamma_\ell \gamma_m^*}{\sum_{i,j \in R(S_k)} |\gamma_i|^2 |\gamma_j|^2} \kb{\ell}{m}.
\end{equation}
We now verify that this matrix is Hermitian and therefore a valid observable. We will use the following:
\begin{align}
    \hat{y}_k &= \hat{y}_{-k}^* \label{eq:dftconjsym},\\
    R(-k) &= \{(m, \ell) \text{ for all } (\ell, m) \in R(k) \} := R(k)^T \label{eq:Rrev}.
\end{align}
Eq.~\eqref{eq:dftconjsym} follows from our assumption that $y_j \in \R \, \forall \, j\in[n]$. And so
\begin{align}
    M_{\ell m} &= \hat{y}_k^* \frac{\gamma_\ell \gamma_m^*}{\sum_{i,j \in R(S_k)} |\gamma_i|^2 |\gamma_j|^2}  \quad \forall \, (\ell, m) \in R(S_k) \\
    \label{eq:3.1} &=\hat{y}_{-k} \frac{\gamma_\ell \gamma_m^*}{\sum_{j,i \in R(S_{-k})^T} |\gamma_i|^2 |\gamma_j|^2}  \quad \forall \, (\ell, m) \in R(S_{-k})^T \\
     \label{eq:3.2} &=\left(\hat{y}_{-k}^* \frac{\gamma_m\gamma_\ell^* }{\sum_{i,j \in R(S_{-k})} |\gamma_i|^2 |\gamma_j|^2}\right)^*  \quad \forall \, (m, \ell) \in R(S_{-k}) \\
    &= M_{m \ell}^*. \label{eq:finalline}
\end{align}
In line~\eqref{eq:3.1} we have used Eq.~\eqref{eq:dftconjsym}, while in line~\eqref{eq:finalline} we have observed that Eq.~\eqref{eq:Mopt_piecewise} holds only with respect to any fixed partition;  $M_{\ell m}$ and $M_{m \ell}$ must be computed with respect to distinct partitions. The optimal model may now be rewritten in terms of base frequencies as
\begin{align}
    f^{opt}(x) &= \sum_{k \in \Omega_n} \hat{y}_k \sum_{\ell=a_k}^{b_k} e^{i 2\pi (k + n\ell)x}    \frac{ \sum_{i, j \in R(k + n\ell)}|\gamma_i|^2 |\gamma_j|^2   }{\sum_{r=a_k}^{b_k} \sum_{c,d \in R(k + nr)} |\gamma_c|^2 |\gamma_d|^2}.
\end{align}
Recall that the optimal classical model derived in Eq.~\eqref{eq:class_solved} is given by 
\begin{equation}
f(x) =  \sum_{k \in \Omega_n} \hat{y}_k \, \sum_{\ell=a_k}^{b_k}  e^{i 2\pi (k + \ell n) x} \frac{\nu_{k + \ell n}}{{\sum_{r =a_k}^{b_k} \nu_{k + nr}}}.
\end{equation}
Then despite Eq.~\eqref{eq:qmodel_2} not having a clear decomposition into scalar \fweight{s} and \tcoef{s}, we can identify the \fweight{s} of the optimized quantum model as
\begin{equation}\label{eq:lambda_comp}
    \nu_{k}^{opt} := \sum_{i, j \in R(k)} |\gamma_i|^2 |\gamma_j|^2,
\end{equation}
which recovers the same form of the optimal classical model of  Eq.~\eqref{eq:class_solved}. This means that the behavior of the \fweight{s} of the (optimal) general quantum model are strongly controlled by the degeneracy sets $R(k)$. The generalization error of the quantum model is also described by Eq.~\eqref{eq:class_err} of the main text under the identification of Eq.~\eqref{eq:lambda_comp}, and therefore exhibits a tradeoff that is predominantly controlled by the degeneracies $R(k)$ of the data-encoding Hamiltonian.

We can now  substitute $\gamma_\ell = 1/\sqrt{d}$ to recover the optimal observable for the simplified model derived in Sec~\ref{sec:toy_model} of the main text using other means, namely
\begin{align}
    M_{ m \ell}^{opt} &= \hat{y}_k \frac{d}{\sum_{r=a_k}^{b_k}  |R(k + nr)|}, \\
    f(x) &= \sum_{k \in \Omega_n} \hat{y}_k \sum_{q=a_k}^{b_k} e^{i 2\pi (k + nq)x}    \frac{|R(k + nq)|}{\sum_{r=a_k}^{b_k}  |R(k + nr)|}, \\
    \nu_{k} &= \frac{|R(k)|}{d^2}.
\end{align}

\subsection{Computing \fweight{s} of typical quantum models}

In Sec~\ref{sec:toy_model} we introduced a simple model with a uniform amplitude state $|\Gamma\rangle$ as input and demonstrated that the \fweight{s} of this simple quantum model are completely determined by the sets $R(k)$ induced by the encoding strategy. We now wish to extend the intuition that the behavior of the optimized general quantum model is strongly influenced by the distribution of the degeneracies $|R(k)|$. We do so by evaluating the optimal quantum models with respect to an ``average'' state preparation unitary $U$. We can compute the average value of $|\gamma_i|^2 |\gamma_j|^2$ for $U$ sampled uniformly from the Haar distribution using standard results from the literature \cite{puchala2011symbolic}\footnote{Note that since $\nu_k^{opt}$ is invariant with respect to $\gamma_i \rightarrow e^{i\phi}\gamma_i$ a spherical measure would suffice here.}:
\begin{align}
    \underset{U \sim \U{d}}{\E} |\gamma_i|^2 |\gamma_j|^2  = \underset{U \sim \U{d}}{\E} U_{i0}U_{i0}^* U_{j0} U_{j0}^* 
    =  \frac{\delta_i^j + 1}{d(d+1)} .
\end{align}
Since $(i, i) \in R(0)$ we can then compute the \fweight{s} of the optimal model according to\footnote{It is implied that we compute the optimal $\nu$ with respect to each distinct $U$ sampled independently and uniformly with respect to the Haar measure -- without optimizing $M$ with respect to each $U$ we would find the trivial result $$\underset{U \sim \U{d}}{\E} f(x) =\underset{U \sim \U{d}}{\E}  \tr{M \rho(x)} = d^{-1}\tr{M}$$.}
\begin{align}\label{eq:haar_nu}
   \underset{U \sim \U{d}}{\E} \nu_k^{opt}= \underset{U \sim \U{d}}{\E}  \sum_{i, j \in R(k)}|\gamma_i|^2 |\gamma_j|^2 = \frac{|R(k)|}{d(d+1)} + \delta_0^k \frac{1}{d+1}.
\end{align}
From Eq.~\eqref{eq:haar_nu} we see that the \fweight{s} of a quantum model optimized with respect to random $U$ are completely determined by the degeneracies $R(k)$. This expected value is useful but does not fully characterize the behavior of an encoding strategy. To demonstrate that this average behavior is meaningful, we would further like to verify that the \fweight{s} corresponding to a random $U$ concentrate around the mean of Eq.~\eqref{eq:haar_nu}. We characterize this by computing the variance
\begin{equation}
    \text{Var}(\nu_k) = \underset{U \sim \U{d}}{\E} \left(\nu_k^2 \right) - \left(\underset{U \sim \U{d}}{\E} \nu_k\right)^2,
\end{equation}
where we have dropped the superscript on $\nu_k^{opt}$ for brevity. This computation requires significantly more counting arguments dependent on the structure of $R(k)$. When $k \neq 0$, we identify cases for which $i\neq j$ whenever $(i, j) \in R(k)$:
\begin{align}
    \nu_k^2 &= \sum_{i, j \in R(k)}\sum_{\ell, m \in R(k) } |\gamma_i|^2 |\gamma_j|^2 |\gamma_\ell|^2 |\gamma_m|^2 
    \\&=  \sum_{i, j \in R(k)}\left( |\gamma_i|^4 |\gamma_j|^4 +  \sum_{\substack{\ell, m \in R(k) \\ \ell \neq j, m = i} } |\gamma_i|^4 |\gamma_j|^2 |\gamma_\ell|^2\right.\nonumber
    \\&\qquad \qquad \qquad \qquad \left. + \sum_{\substack{m, \ell \in R(k) \\ \ell = j, m \neq i } } |\gamma_i|^2 |\gamma_j|^4 |\gamma_m|^2 + \sum_{\substack{m, \ell \in R(k) \\ \ell \neq j, m \neq i } } |\gamma_i|^2 |\gamma_j|^2 |\gamma_\ell|^2 |\gamma_m|^2 \right).\label{eq:broken_sum}
\end{align}
The expected values for of these terms are evaluated using  the observation that the vector $(|\gamma_1|^2, \dots, |\gamma_s|^2)$ is distributed uniformly on the $d$-simplex leading to simple expressions for the following expected values \cite{puchala2011symbolic}:
\begin{align}
    \underset{U \sim \U{d}}{\E}|\gamma_p|^4 |\gamma_q|^4 &=  4/D, \\
    \underset{U \sim \U{d}}{\E}|\gamma_p|^4 |\gamma_q|^2 |\gamma_r|^2  &= 2/D, \\    
    \underset{U \sim \U{d}}{\E}|\gamma_p|^2 |\gamma_q|^2 |\gamma_r|^2 |\gamma_s|^2  &= 1/D, 
\end{align}
where $p,q,r,s = 1, \dots, d$ and $p \neq q \neq r \neq s$, and $D := (d+3)(d+2)(d+1)d$. In computing the expected value of Eq.~\eqref{eq:broken_sum}, by linearity the terms of each sum will become a constant and only the number of items in each sum will be relevant. The first sum of Eq.~\eqref{eq:broken_sum} contains $|R(k)|$ many terms and the total number of terms is $|R(k)|^2$, and so we only need compute the number of elements in the two middle sums of Eq.~\eqref{eq:broken_sum} (which contain an equal number of terms due to the symmetry $R(k) = R(-k)^T$). These computations may be carried out by brute-force combinatorics and are summarized in Table~\ref{tab:degen_tedious} for a few of the models studied in this work.

\begin{table}
    \centering
    \begin{tabular}{|c|c|c|}
    \hline
        Encoding strategy &  $|\{(i,j,\ell,m): m\neq i, \ell= j, (i,j), (\ell,m) \in R(k)\}|$  & $\text{Var}(\nu_k^{opt})$  \\
        \hline
       Hamming   & $\sum_{p=0}^{n_q-2k} \begin{psmallmatrix} n_q \\ p\end{psmallmatrix}\begin{psmallmatrix} n_q \\ p+|k|\end{psmallmatrix}\begin{psmallmatrix} n_q \\ p+2|k|\end{psmallmatrix}$ & - \\
       Binary  &  $\max(2^{n_q} - 2|k|, 0)$ & $O\left( 2^{-3n_q} \right)$ \\
        Golomb  & 0 &  $\mathcal{O}(d^{-4})$ \\
       \hline
    \end{tabular}
    \caption{The size of the subset of $R(k)\times R(k)$ with a single repeated index computed for different encoding strategies and the corresponding scaling of the variance of \fweight{s} (all entries assume $k\neq 0$). The Hamming encoding strategy computation works as follows: Each bitstring $i$ with weight $p$ (of which there are $n$-choose-$p$) is paired with $n$-choose-$(p+k)$ many bitstrings $j$ with weight $p+k$. Then, taking $\ell=j$ there will be $n$-choose-$(p+2k)$ many bitstrings $m$ with weight $(p+k) + k = p + 2k$. Summing over all such $p$'s where $p+2k \leq n_q$ yields the desired result, however this computation does not admit a clear closed-form expression and so we have omitted the corresponding scaling of $\text{Var}(\nu_k^{opt})$ for the Binary encoding strategy.}
    \label{tab:degen_tedious}
\end{table}

\begin{figure}
    \centering
    \includegraphics[width=.8\textwidth]{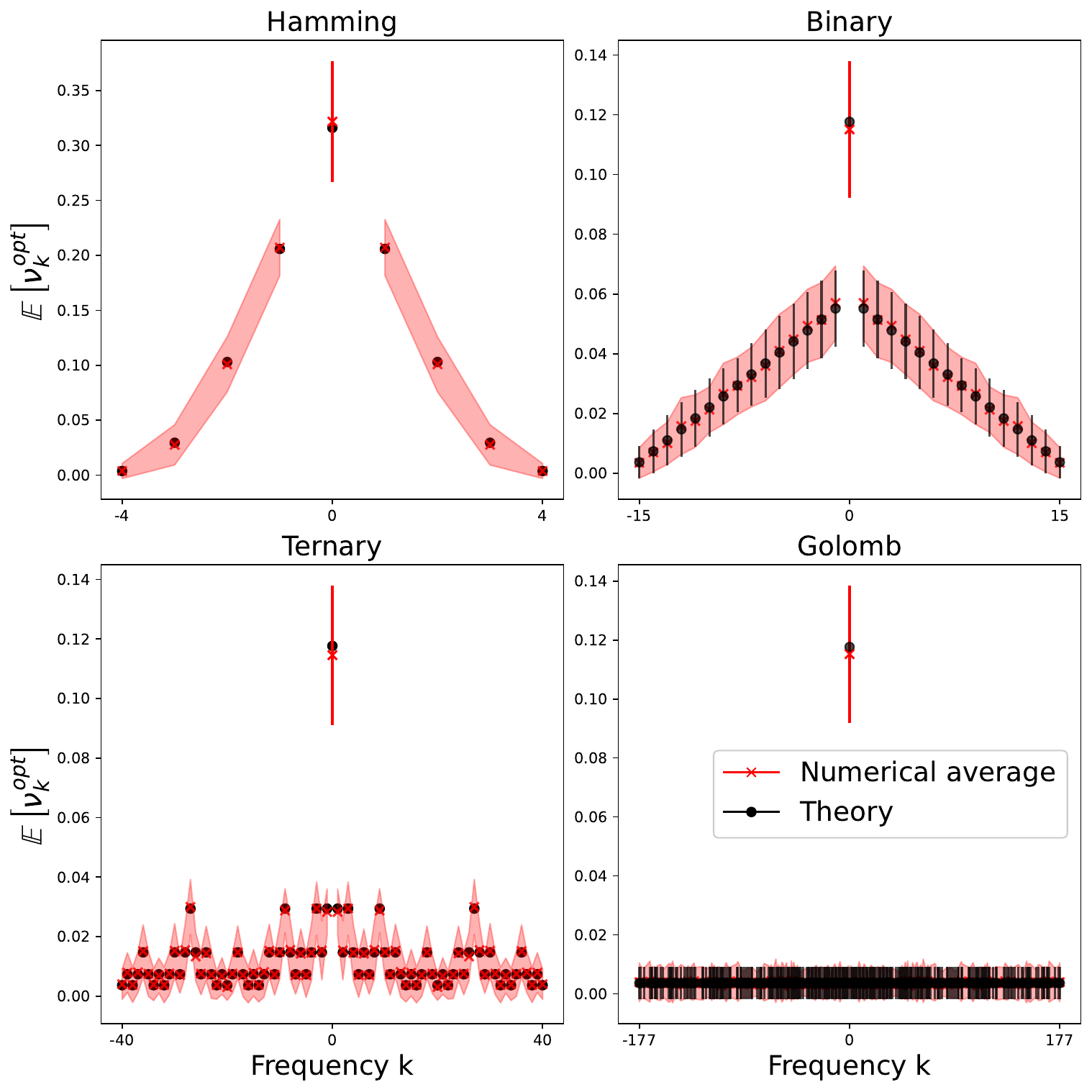}
    \caption{Plotting $\E \left[\nu_k^{opt}\right]$ and $\text{Var}\left[\nu_k^{opt}\right]^{1/2}$ demonstrates tight concentration of the distribution of $\nu_k^{opt}$ around its mean with respect to Haar-random $U$ in the general quantum model. For the Binary and Golomb encoding strategies we plot Eqs.~\eqref{eq:binary_nu_var} and \eqref{eq:golomb_nu_var} respectively. For all models we also include averages and variances computed by sampling states uniformly from the Haar measure numerically.}
    \label{fig:nu_plots}
\end{figure}

For the Binary encoding strategy we compute
\begin{align}
    \E (\nu_k^2) = \begin{cases}
    \frac{5d - 7k + (d-k)^2 }{D}, & 0< 2k \leq d \\
    \frac{3(d-k) + (d-k)^2}{D}, & 2k > d,
    \end{cases}
\end{align}
and so
\begin{align}\label{eq:binary_nu_var}
    \text{Var}(\nu_k) = 
    \left\{\!\begin{aligned}
    &\frac{(5d - 7k) d (d+1) - (d-k)^2(4d + 6) }{Dd(d+1)} \text{ if } 0 <2k \leq d \\
    &\frac{3(d-k) d (d+1) - (d-k)^2(4d + 6) }{Dd(d+1)} \text{ if } 2k > d
\end{aligned}\right\} \leq \mathcal{O}(d^{-3}) .
\end{align}
Importantly, $\E[\nu_k] \geq \mathcal{O}(d^{-2})$ for the Binary encoding strategy. Taking $d=2^{n_q}$ corresponding to $n_q$ qubits, then while the mean decays exponentially in the number of qubits $n_q$, the variance decays exponentially faster. For the Golomb encoding strategy, the calculation is comparatively straightforward, yielding for $k \neq 0$
\begin{equation}\label{eq:golomb_nu_var}
    \text{Var}(\nu_k) = \frac{3d^2 - d - 6}{D d (d+1)} = \mathcal{O}(d^{-4}),
\end{equation}
with the variance again decaying significantly faster in $d$ than the mean. Figure~\ref{fig:nu_plots} shows the average and variance of $\nu_k^{opt}$ for the general quantum model with $U$ sampled uniformly with respect to the Haar measure. We find tight concentration of $\nu_k^{opt}$ around its average in each of these cases

Why is it that the \fweight{s} $\nu_k^{opt}$ of the quantum model given by Eq.~\eqref{eq:lambda_comp} only appear after deriving the optimal observable? One explanation is that while the classical Fourier Features model utilizes random Fourier features such that components of $\phi(x)$  are mutually orthogonal on $[0, 1]$, the quantum model does not.  Consider the operator
\begin{equation}
   \Sigma := \int_\X \Vecc{\rho(x)} \Vecc{\rho(x)}^\dagger p(x) dx = \int_\X \rho(x) \otimes \rho^T(x) p(x) dx,
\end{equation}

where $p: \X \rightarrow \R$ is the probability density function describing the distribution of data in $\X$ and $\text{Vec}: \C^{d \times d} \rightarrow \C^{d^2}$ is the \textit{vectorization map} that acts by stacking transposed rows of a square matrix into a column vector. $\Sigma$ is analogous to a classical covariance matrix, and here determines the correlation between components of $\rho(x)$ (with the second equality holding when $\rho(x)$ is a pure state for all $x$). From Eq.~\eqref{eq:weighted_ff} describing the classical Fourier features it is straightforward to compute $\E_{x\sim [0,1]} [\phi \phi^\dagger]  = \I_d$, demonstrating that the classical Fourier features are indeed orthonormal. However, under the identification of the feature vector $\phi(x) \rightarrow \Vecc{\rho(x)}$ in the quantum case, the same is not true for the quantum model:
\begin{align}
    \Sigma &= \int_0^1 \left(\sum_{k\in\Omega} e^{i2\pi kx} \sum_{i, j \in R(k)} \gamma_i \gamma_j^* |ij\rangle \right) \left(\sum_{k'\in\Omega} e^{i2\pi k'x} \sum_{\ell, m \in R(k')} \gamma_\ell^* \gamma_m \langle \ell m | \right) dx
    \\&= \sum_k \sum_{\substack{i, j \in R(k) \\ \ell,m\in R(k)}} \gamma_i \gamma_j^* \gamma_\ell^* \gamma_m |ij\rangle\langle \ell m|.
\end{align}

Thus $\Sigma$ is not diagonal in general, as many components of $S(x)|\Gamma\rangle$ each contribute to a single frequency $k$. Nor will it be possible in general to construct a quantum feature map via unitary operations that does act as an orthogonal Fourier features vector. As $\Sigma$ is positive semidefinite by construction, there exists a spectral decomposition $\Sigma = W D W^\dagger$ with diagonal $D$ and $d^2 \times d^2$ unitary $W$. However the linear operator $\Phi \in L(d)$ acting on $d$-dimensional states according to $\Vecc{\Phi(\rho)} = W^\dagger \Vecc{\rho(x)}$ will not be unitary in general, and thus it may not be possible to prepare $\rho$ in such a way that the elements of $\Vecc{\rho(x)}$ consist of orthonormal Fourier features.

\subsection{The effect of state preparation on \fweight{s} in the general quantum model}

We now discuss how the state preparation unitary $U$ may affect the \fweight{s} $\nu_k^{opt}$ in a quantum model, and what choices one can make to construct a $U$ that gives rise to a specific distribution of \fweight{s} $\nu_k^{opt}$.

As an example, we consider the general quantum model using the Hamming encoding strategy and an input state $|\Gamma\rangle$. Computing the \fweight{s} $\nu_k^{opt}$ for the Hamming encoding strategy depends on the amplitudes $\gamma_i$ and $\gamma_j$ for which the weights of the indices $i, j$ differ by $k$. We have
\begin{align}
    \nu_k^{opt} = \sum_{i, j \in R(k)} |\gamma_i|^2 |\gamma_j|^2 = \sum_{\substack{i,j: \\ w(j) - w(i) = k}} |\gamma_i|^2 |\gamma_j|^2.
\end{align}
We will now show how the \fweight{} $\nu_k^{opt}$ may be computed with respect to a rebalanced input state which distributes each amplitude $\gamma_i$ among all computational basis states with index having weight $w(i)$. We define this rebalanced state on $n$ qubits as
\begin{align}
    |\Gamma'\rangle &= \sum_{i\in\{0,1\}^n} \gamma_i ' |i\rangle = \sum_{i=0}^n \sum_{j \in W(i)} \phi_i |j\rangle\label{eq:rebalanced_state},\\
    \phi_i &= \frac{1}{\sqrt{|W(i)|}} \left(\sum_{j \in W(i)} |\gamma_j|^2\right)^{1/2} \label{eq:rebalanced_coeff}.
\end{align}
where $W(i) = \{j: j \in \{0,1\}^n, w(j) = i\}$ is the set of weight-$i$ indices and $|W(i)| = \binom{n}{i}$. Observing that the amplitudes of $|\Gamma'\rangle$ satisfy $|\gamma_i'|^2= \phi_{w(i)}^2$, we can derive for $k\geq 0$
\begin{align}
    \nu_k^{opt} &= \sum_{\substack{i,j: \\ w(j) - w(i) = k}} |\gamma_i|^2 |\gamma_j|^2 
    \\&= \sum_{\ell=0}^{n-k} \left( \sum_{i: w(i) = \ell} |\gamma_i|^2 \right) \left( \sum_{j: w(j) = \ell + k} |\gamma_j|^2 \right)
    \\&= \sum_{\ell=0}^{n-k} \left( |W(\ell)| \frac{\sum_{i: w(i) = \ell} |\gamma_i|^2 }{|W(\ell)|} \right)  \left( |W(\ell + k)|\frac{\sum_{j: w(j) = \ell + k} |\gamma_j|^2}{|W(\ell + k)|} \right)
    \\&= \sum_{\ell=0}^{n-k} |W(\ell)| \phi_\ell^2 |W(\ell + k)| \phi_{\ell + k}^2
    \\&= \sum_{\ell=0}^{n-k} \left(\sum_{i: w(i) = \ell} \phi_{w(i)}^2\right)\left( \sum_{j: w(j) = \ell + k} \phi_{w(j)}^2 \right)
    \\&= \sum_{\substack{i,j: \\ w(j) - w(i) = k}} |\gamma_i'|^2 |\gamma_j'|^2 .
\end{align}
Therefore, the \fweight{s} computed from $|\Gamma\rangle = U|0\rangle$ and the rebalanced state $|\Gamma'\rangle$ are identical. We can emphasize the significance of this observation by rewriting $|\Gamma ' \rangle$ of Eq.~\eqref{eq:rebalanced_state} as
\begin{align}
    |\Gamma'\rangle &= \sum_{i=0}^n \phi_i |\Phi_i\rangle, \label{eq:rewrite} 
\end{align}
where $|\Phi_i\rangle = \sum_{j \in W(i)} |j\rangle$ is an (unnormalized) superposition of bitstrings with weight $i$. The state $|\Gamma'\rangle$ of Eq.~\eqref{eq:rewrite} has only $n+1$ real parameters $\phi_i$, and is invariant under operations that are restricted to act within the subspace spanned by components of $|\Phi\rangle$. This invariance greatly reduces the class of unitaries $U$ that affect the distribution of \fweight{s} $\nu_k^{opt}$ and enables some degree of tuning for these parameters.

\subsubsection{Vanishing gradients in preparing \fweight{s}} 

We briefly remark on the possibility of training \fweight{s} $\nu_k^{opt}$ of the general quantum model variationally. For example, one could consider defining a desired distribution of \fweight{s} as $\phi_k \in \R^{|\Omega|}$ and then attempting to tune parameters of $U$ in order to minimize a cost function such as\footnote{As $\nu_k$ represents a probability distribution, other distance measures such as relative entropy or earth-mover's distance may be considered more appropriate. However using alternative cost functions will not affect the main arguments here.}
\begin{equation}\label{eq:loss}
    L = \sum_{k \in \Omega} |\nu_k^{opt} - \phi_k|^2.
\end{equation}
We will now demonstrate that for a sufficiently expressive class of state preparation unitaries $U$, such an optimization problem will be difficult on average. For simplicity, we follow the original formulation of barren plateaus presented in Ref.~\cite{mcclean2018barren}: Let $U$ be defined with respect to a set of parameters $\boldsymbol{\theta} \in \R^L$ as
\begin{equation}
    U = \prod_{\ell=1}^L U(\theta_\ell) W_\ell
\end{equation}
with $U(\theta_\ell) = \exp(i \theta_\ell V_\ell)$, $V_\ell$ being a $d$-dimensional Hermitian operator, and $W_\ell$ being a $d$-dimensional unitary operator. Pick a parameter $\theta_m$ and define $U_+ = \prod_{\ell=m}^L U(\theta_\ell) W_\ell$ and $U_- = \prod_{\ell=1}^m U(\theta_\ell) W_\ell$, and observe that 
\begin{equation}
    \partial_{\theta_m} |\gamma_j|^2 = i\langle 0| U_-^\dagger [V_m, U_+^\dagger \dm{j} U_+] U_- |0\rangle.
\end{equation}
We can compute the derivative of $\nu_k$ with respect to $\theta_m$ using the chain rule:
\begin{align}
    \frac{\partial \nu_k}{\partial \theta_m} &= \sum_{j=1}^d \frac{\partial \nu_k}{\partial |\gamma_j|^2}\frac{\partial |\gamma_j|^2}{\partial \theta_m}
    \\&= \sum_{j=1}^d \left( \sum_{a, b \in R(k)} \left(|\gamma_a|^2 \delta_b^j +\delta_a^j|\gamma_b|^2 \right) \right) i\langle 0| U_-^\dagger [V_m, U_+^\dagger \dm{j} U_+] U_- |0\rangle
    \\&= i\sum_{a, b \in R(k)} \left[\tr{\rho_- H_a}  \tr{\rho_-[V_m, H_b]}  + \tr{\rho_- [V_m, H_a]} \tr{\rho_- H_b}   \right] \label{eq:partial_m}
\end{align}
where $\rho_- = U_-\dm{0} U_-^\dagger$ and $H_x = U_+^\dagger \dm{x} U_+$, and we used the equality 
\begin{align}
 |\gamma_a|^2  =  \langle 0| U_-^\dagger U_+^\dagger \dm{a}U_+ U_-|0\rangle = \tr{\rho_- H_a}.
\end{align}

We now show that each term in this sum vanishes by following Ref.~\cite{mcclean2018barren} in letting $U_-$ be sampled from a distribution that forms a unitary 2-design: 
\begin{align}
    \E_{U_- \sim U(d)} \tr{\rho_- H_a}&  \tr{\rho_-[V_m, H_b]}  
    \\&= \tr{ \E_{U_- \sim U(d)} \left[\rho_- \otimes \rho_- \right] H_a \otimes [V_m, H_b] }
    \\&=\frac{1}{d(d+1)} \tr{ \left(\I_d + \sum_{i,j=1}^d |ij\rangle \langle ji| \right) H_a \otimes [V_m, H_b] } \label{line:123}
    \\&= \frac{ \tr{ [V_m, H_b] H_a}}{d(d+1)}
    \\&=\frac{ V_m U_+^\dagger \dm{b} U_+ U_+^\dagger \dm{a} U_+ - U_+^\dagger \dm{b} U_+ V_m U_+^\dagger \dm{a} U_+ }{d(d+1)}  
    \\&= \frac{\delta_a^b \langle a| U_+ V_m U_+^\dagger |b\rangle - \delta_a^b \langle b| U_+ V_m U_+^\dagger |a\rangle }{d(d+1)} 
    \\&= 0,
\end{align}
where in line~\eqref{line:123} we have used a common expression for $\E_{U\sim U(d)} [\rho \otimes \rho]$ in terms of the projector onto the symmetric subspace (e.g. \cite{Roberts_2017}). Substituting this result into Eq.~\eqref{eq:partial_m} we find
\begin{equation}
    \E_{U \sim U(d)} \left(\frac{\partial \nu_k}{\partial \theta_m}  \right) = 0
\end{equation}
By extension, the gradient of a loss function of the form of Eq.~\eqref{eq:loss} will vanish for expressive enough state-preparation unitaries $U$, suggesting that solving for a choice of $U$ to induce a specific distribution of \fweight{s} $\nu_k^{opt}$ will be infeasible in practice.

\section{Determining the degeneracy of quantum models}\label{app:combinatorics}

Here we will develop a theoretical framework for manipulating the degeneracy (and therefore \fweight{s}) of quantum models. We will begin with choosing the data-encoding operator $S(z) = \exp(-i H z)$ to be separable, generated from a Hamiltonian of the form 
\begin{equation}\label{eq:Ham_model}
    H = \mathbf{r} \cdot \mathbf{Z} = \sum_{j=0}^{n-1} r_j Z^{(j)},
\end{equation}
and $Z^{(j)}$ denotes the operator that applies a Pauli $Z$ operator to the $j^{th}$ register and acts trivially elsewhere, and $\mathbf{r} \in \R^n$.  One can show that the diagonal elements of $H$ are given by
\begin{equation}\label{eq:sep_spec}
    \lambda_j = H_{jj} = 2 (\mathbf{r} \cdot \mathbf{j}) - \norm{\mathbf{r}}_1,
\end{equation}
where here and elsewhere we use interchangeably a bitstring denoted as $j = j_0 j_1 \dots j_{n-1}$ with decimal value $j = \sum_k j_k 2^j\in [0, 2^n - 1]$, and $\mathbf{j} \in \{0,1\}^n$ as its corresponding vector representation. The Fourier spectrum and degeneracy of the encoding strategy are then computed as
\begin{align}
    \Omega &= \{ \lambda_j - \lambda_i: i,j = 0, \dots, 2^n-1\} \label{eq:sep_omega}
    \\&= \{ 2 (\mathbf{r} \cdot (\mathbf{j} - \mathbf{i})): \mathbf{i}, \mathbf{j} \in \{0, 1\}^n\},
    \\ R(\omega) &= \{(i,j): 2\mathbf{r} \cdot (\mathbf{j} - \mathbf{i}) = k \}.
\end{align}
Note that the subtraction in the definition of $\Omega$ does not preserve $\Z_2$, i.e. $(\mathbf{j} - \mathbf{i}) \in \{-1, 0, 1\}^n$.  From Eq.~\eqref{eq:sep_omega} we see that the largest possible size for $|\Omega|$ is $3^n$, the number of unique choices for $(\mathbf{j} - \mathbf{i})$. Any encoding strategy of Eq.~\eqref{eq:Ham_model} therefore introduces a \textit{combinatorial degeneracy}, since the set $\{(\mathbf{j} - \mathbf{i}): \mathbf{i}, \mathbf{j} \in \{0, 1\}^n\}$ is the image of a surjective map on $\{0, 1\}^n \times \{0, 1\}^n$, with each $(\mathbf{j} - \mathbf{i})$ occurring with multiplicity 
    \begin{equation}\label{eq:combdegen}
        \#(\mathbf{j} - \mathbf{i}) = 2^{n - \norm{(\mathbf{j} - \mathbf{i})}_1}.
    \end{equation}
    
To prove this, construct the set $\{(\mathbf{j} - \mathbf{i}): \mathbf{i}, \mathbf{j} \in \{0, 1\}^n\}$ by reflecting the hypercube $\{0,1\}^n$ over $2^n$ distinct axes, and count the number of vertices shared by among reflected images -- this gives the desired degeneracy. Many choices of $\mathbf{r}$ will reduce $|\Omega|$ by reducing the size of the image of the map $(\mathbf{i}, \mathbf{j}) \rightarrow \mathbf{r} \cdot (\mathbf{j} - \mathbf{i})$. Conversely, the spectrum $\Omega$ will saturate $|\Omega| = 3^n$ only under particular conditions on $\mathbf{r}$. For example with $n=2$, $|\Omega|=9$ is achieved only if $\mathbf{r}$ satisfies all of the following conditions
    \begin{align}\label{eq:limits}
        r_0 \neq 0, \qquad r_0 \neq \pm \frac{1}{2} r_1, \qquad r_0 \neq \pm r_1, \qquad r_0 \neq \pm 2 r_1, \qquad 
    \end{align}
i.e., the inverse of the dot product $\mathbf{r} \cdot \mathbf{d}$ with respect to an input $\mathbf{d}$ exists if the elements of $\mathbf{r}$ satisfy $r_j \neq r_i \text{ mod } 2^p$ for some integer $p$; a more general, recursive statement of this property is provided in Ref.~\cite{shin2022}. Requirements of this form may be proven using the \textit{frame analysis operator} $T_n \in \Z^{3^n \times n}$, the rows of which contain each element of $\{-1, 0, 1\}^n$. And so to demonstrate Eq.~\eqref{eq:limits} above, if we let $\mathbf{k}\in \R^{3^n}$ be a vector containing (repeated) frequencies in $\Omega$, we observe that $T_2 \mathbf{r} = \mathbf{k} \in \R^9$ will have unique entries only under specific conditions on $\mathbf{r}$. $\Omega$ being the result of a frame reconstruction of $\mathbf{r}$ also provides a direct way to tune the elements of $\Omega$: given a length $3^n$ vector $\mathbf{k}$ containing the (not necessarily unique) desired elements of $\Omega$, one can then compute $H$ by solving $\mathbf{r} = T_n^\dagger \mathbf{k}$.

\subsection{Hamming encoding strategy}\label{sec:hamming_k}

We instantiate the Hamiltonian of Eq.~\eqref{eq:Ham_model} for 
$$
\mathbf{r} = \frac{1}{2}(1, 1, \dots, 1) \in \R^n.
$$
This introduces many degeneracies into the spectrum; using $T_n \mathbf{r} = \mathbf{k}$ we immediately recover the result of Ref.~\cite{schuld2021} that the unique elements of $\mathbf{k}$ are given by
\begin{equation}
    \Omega = \{ -n, -(n-1), \dots, 0, \dots, n-1, n\},
\end{equation}
with $|\Omega| = 2n + 1$. To recover the degeneracy of each frequency we compute
\begin{align}
    \lambda_i - \lambda_j = \frac{1}{2}\sum_{t=1}^n  (-1)^{i_t} - \frac{1}{2}\sum_{t=1}^n (-1)^{j_t} = \frac{1}{2}\sum_{t=1}^n  (1 - 2 i_t) - \frac{1}{2}\sum_{t=1}^n (1 - 2 j_t) = w(j) - w(i),
\end{align}
where $w: \{0,1\}^n \rightarrow [n]$ is the the weight of a bitstring $b = b_{n-1} \dots b_1 b_0 \in \{0, 1\}^n$ defined as
\begin{equation}
    w(b) = \sum_{k=1}^n b_k,
\end{equation}
and when $b$ is written as an integer it should be interpreted according to its binary value. It will be useful to consider subsets of bitstrings having a constant weight. For each $k \in [n]$ (results for $k < 0$ follow by symmetry), we now define the $k$-weight  set $W(k)$ as
\begin{equation}\label{eq:Wdef}
    W(k) = \{j \in \{0,1\}^n: w(j) = k \}.
\end{equation}
 Then the degeneracy set is exactly the set of bitstring indices differing by constant weight:
\begin{align}
    R(k) &= \{(i, j) \in \{0,1\}^n \times \{0,1\}^n: w(j) - w(i) = k \}
    \\&= \bigcup_{p=k}^n W(p) \times W(p- k).
\end{align}
where $\times$ denotes the Cartesian product (this motivates the name ``Hamming encoding strategy'', as the frequencies and degeneracy of this model arise from index pairs having fixed Hamming weight). From counting arguments, the degeneracy of this model is then 
\begin{equation}
     |R(k) | = \sum_{j=1}^{n-k} \begin{pmatrix}
    n \\ j
    \end{pmatrix}\begin{pmatrix}
    n \\ k +j
    \end{pmatrix} = \begin{pmatrix}
    2n \\ n - k \end{pmatrix}.
\end{equation}

\subsection{Binary encoding strategy}\label{sec:binary_k}
We instantiate the Hamiltonian of Eq.~\eqref{eq:Ham_model} for 
$$
\mathbf{r} = \frac{1}{2}(2^{n-1}, \dots, 4, 2, 1) \in \R^n,
$$
such that $\mathbf{r}$ generates the decimal value of binary vector. We know from relations like those given in Eq.~\eqref{eq:limits} that $|\Omega| << 3^n$, and indeed using Eq.~\eqref{eq:sep_spec} we find that $\lambda_j = (j - (2^n - 1)) / 2$ resulting in a frequency spectrum of
\begin{equation}
    \Omega = \{ -2^n+1, -2^n + 2, \dots, 0, 1, \dots 2^n - 1\},
\end{equation}
such that $|\Omega| = 2^{n+1} - 1$, and the degenerate index set of the kernel for frequency $k$ is equivalent (up to permutations) to the indices corresponding to nonzero elements of an elementary Toeplitz matrix with 1's on the $k$-th diagonal:
\begin{equation}
    R(k) = \{(i, j): i = j - k\,\,  i, j\in [2^n]  \}.
\end{equation}

\subsection{Ternary encoding strategy}\label{sec:ternary_k}
We instantiate the Hamiltonian of Eq.~\eqref{eq:Ham_model} for 
$$
\mathbf{r} = \frac{1}{2}(3^{n-1}, \dots, 9, 3, 1) \in \R^n,
$$
resulting in the spectrum
\begin{equation}
    \Omega = \Bigl\{- \frac{1}{2} (3^n - 1) ,- \frac{1}{2} (3^n - 1) +1, \dots, 0, \dots, \frac{1}{2} (3^n - 1) \Bigr\}.
\end{equation}
This spectrum (also studied in Ref.~\cite{shin2022}) is interesting to study because it is the unique choice giving rise to a Fourier spectrum with frequencies spaced by $1$ saturating the $|\Omega| = 3^n$ limit for spectra generated by separable Hamiltonians. Given $k =2 \mathbf{r} \cdot \mathbf{d}$, then $\mathbf{d} = \mathbf{i} - \mathbf{j}$ may be uniquely recovered from $k$.  Suppose $k$ has a ternary representation
\begin{equation}\label{eq:ternary_def}
    k = k_{n}\dots k_1 k_0 = \sum_j k_j 3^j,
\end{equation}
with $n = \lceil \log_3 k \rceil$ and $k_j \in \{0, 1, 2\}$. Then defining the operation $t: \Z \rightarrow \{0,1,2\}^n$ which converts decimals to their vector-form ternary representation according to $t(k) = (k_0, k_1, \dots, k_{n-1})$, then the degeneracy of this encoding strategy follows directly from the combinatorial degeneracy given in Eq.~\eqref{eq:combdegen}:
\begin{align}
    |R(k)| = 2^{n - \norm{T(k)}_1},
\end{align}
where the shifted ternary operation is defined for $k \in [0, 3^n - 1]$
\begin{align}\label{eq:shift_ternary}
    T(k) = t\left(k + \norm{\mathbf{r}}_1\right) - \mathbf{1},
\end{align}
and $\mathbf{1} = (1, 1, \dots, 1) \in \Z^n$ is the array of all ones. Intuitively, this computation converts a signed ternary expression $k=\sum_j k_j 3^j$ (with $\mathbf{k} \in \{-1, 0, 1\}$) to an unsigned ternary expression provided in Eq.~\eqref{eq:ternary_def}.

\subsection{Nonseparable models}

Given a data-encoding Hamiltonian $H = \diag{\Lambda}$ with $\Lambda \in \R^d$, a key feature of the associated spectrum is that it is preserved under permutations of $\Lambda$. This means that a large class of models where $S(x)$ is generated by a separable $H = H_0 \otimes H_1 \otimes H_{n-1}$ are capable of producing the same spectrum as models where $S(x)$ is generated by an nonseparable Hamiltonian. For instance, letting $\sim$ denote equivalence up to permutation and additive shift, it holds that $I \otimes Z \sim Z \otimes Z$, so that a data encoding unitaries
\begin{align}
    S_1(x) &= \exp(i 2\pi r(I \otimes Z) x), \\
    S_2(x) &= \exp(i 2 \pi r (Z \otimes Z) x),
\end{align} 
have identical spectrum and degeneracy properties (differing only in their corresponding degeneracy index sets $R$). Understanding the classes of nonseparable data-encoding Hamiltonians that are equivalent to a separable encoding strategy up to permutation and translation may be an interesting problem for future work.

We therefore limit our discussion of nonseparable models to a particular choice of Hamiltonian which is provably inequivalent to any separable Hamiltonian, as it saturates the largest possible $|\Omega|$ (and smallest values of $|R(k)|$) available to any single-layer quantum model. This can be achieved by setting the diagonal of the data-encoding Hamiltonian as the elements of a Golomb ruler \cite{piccard1939,babcock1953}. The resulting spectrum is nondegenerate for all nonzero frequencies, though it is straightforward to prove that one cannot achieve uniform spacing in the spectrum (i.e., a \textit{Perfect Golomb ruler}) for $d \geq 5$. As a result, the corresponding spectrum of this model generally exhibits gaps between frequencies. Further exploration of the connections to concepts from radio engineering \cite{aktinson1986, robinson1967} and classical coding theory \cite{fang1977} may enrich investigations into the spectral properties of quantum models.





\section{Demonstration of benign overfitting in the general quantum model}\label{app:quantum_bo}

In this section we demonstrate an example of benign overfitting in the general quantum model by explicitly constructing a sequence of state preparation unitaries $U$ for a particular choice of data-encoding unitary $S(d)$. We will consider a   $d$-dimensional version of the Binary encoding strategy constructed from a data-encoding Hamiltonian $H = \diag{0, 1, 2, \dots, d-1}$ that gives rise to
\begin{align}
    \Omega &= \Bigl\{ \frac{-d}{2} + 1, \dots, 0, \dots, \frac{d}{2} - 1 \Bigr\}, \\
    |R(k)| &= d - |k|,
\end{align}
for $k \in \Omega$, and we will require $d$ to be even for simplicity. Suppose the target function spectrum has size $n_0 < n < d$ for some integer $n$ and satisfies $(n_0 + 1) \text{ mod } 4 = 0$. Define the constants
\begin{align}
    c_1 &= \frac{d}{2} - \frac{n_0 + 1}{4}, \\
    c_2 &= \frac{d}{2} + \frac{n_0 + 1}{4}.
\end{align}
We then choose a constant $a \in [0, 2/(n_0 + 1)]$ and prepare $|\Gamma\rangle$ with elements given by
\begin{align}\label{eq:gamma_hat}
    \gamma_j = \begin{cases}
    \sqrt{a}, & j \in [c_1, c_2), \\
    \sqrt{b}, & j \in [0, c_1) \cup [c_2, d),
    \end{cases}
\end{align}
where normalization requires that
\begin{equation}
    b = \frac{ 1 - \frac{n_0 + 1}{2} a}{d - \frac{n_0 + 1}{2}}.
\end{equation}
For this encoding strategy, the optimal \fweight{s} of the quantum model corresponding to positive frequencies $k \in \Omega_+$ are given by \begin{equation}
    \nu_k = \sum_{j=0}^{d - k - 1} |\gamma_j|^2 |\gamma_{j+k}|^2.
\end{equation}
Counting arguments and algebraic simplification lead to
\begin{equation}\label{eq:nu_bo}
    \nu_k = \begin{cases}
        (d - B_0 - 2k) b^2 + 2kab + (B_0 - k) a^2, & 0 \leq k < c_2 - c_1, \\
        (d - 2B_0 - k) b^2 + 2B_0 ab, & c_2 - c_1 \leq k < c_1, \\
        (k - B_0)b^2 + (d - B_0 - 2k) ab, & c_1 \leq k < c_2, \\
        (d-k)b^2, & c_2 \leq k < d,
    \end{cases}
\end{equation}
where $B_0 = (n_0 + 1)/2$ and $\nu_{k+1} \leq \nu_k$ for all $k > 0$. Noting that $b = \mathcal{O}(d^{-1})$ and $a$ is constant, we find that $\nu_k$ has three distinct scaling regimes: $\nu_k = \Omega(1)$ for $k \in \Omega_{n_0}$, $\nu_k = \Theta\left(d^{-2}\right)$ when $k \in [c_2, d)$, and $\nu_k = \mathcal{O}(d^{-1})$ otherwise. We can use this behavior to bound the generalization error of the quantum model in a manner analogous to Sec.~\ref{sec:simple_bo}. Defining $m = |S(k)| - 1$, we bound the bias according to
\begin{align}
    \textsc{bias}^2 &= \sum_{k\in\Omega_{n_0}}  |\hat{g}_k|^2 \frac{\left(\sum_{\ell \in S(k)\backslash k } \nu_\ell \right)^{2}  + \sum_{\ell \in S(k)\backslash k} \nu_\ell^2}{\left(\sum_{\ell \in S(k) } \nu_\ell \right)^{2}}
    \\&\leq \sum_{k\in\Omega_{n_0}}  |\hat{g}_k|^2 \frac{m(m+1) (db^2 + 2B_0 ab)^2}{(a^2 + m b^2)^2} \label{line:240}
    \\&\leq  \left[m (m+1)b^2\right]\left(\frac{db + 2B_0 a}{a^2 + m b^2 } \right)^2 P\label{line:240b} 
    \\&= \mathcal{O} \left( \frac{1}{n^2}\right), \label{eq:qbias_scaling}
\end{align}
where we have defined $P :=  \sum_{k\in\Omega_{n_0}}  |\hat{g}_k|^2$. In line~\eqref{line:240} we have used the monotonicity of $\nu_k$ to bound the components of $S(k) \backslash k$ in the numerator according to the upper bound for the interval $k \in [c_2 - c_1, c_1)$ and applied a lower bound for elements of $S(k) \backslash k$ in the denominator using the interval $k \in [c_2, d)$. The scaling of Eq.~\eqref{eq:qbias_scaling} follows from $m = \mathcal{O}(d/n)$, which implies that $m (m+1)b^2 = \mathcal{O}(n^{-2})$, while the remaining terms in line~\eqref{line:240b} scale as no greater than a constant. To bound the variance, we observe
\begin{align}
   &\sum_{k \in \Omega_{n_0}} \frac{\sum_{\ell \in S(k)} \nu_\ell^2}{\left(\sum_{\ell \in S(k) } \nu_\ell \right)^{2}} 
   \\&\leq \sum_{k \in \Omega_{n_0}} \frac{\left[a^2(B_0 - k) + 2 k ab + (d - B_0 - 2k) b^2\right]^2 + \left[(d - 2B_0 - k) b^2 + 2B_0 ab\right]^2}{(a^2 + mb^2)^2} \label{line:varub1}
   \\&\leq n_0 \frac{(a^2 B_0 + n_0 ab + db^2)^2 + m(db^2 + 2B_0ab)^2}{(a^2 + mb^2)^2}\label{line:varub2}
   \\&= \mathcal{O}(1), \label{line:varub2.5}
\end{align}
which follows identically to the arguments used to reach Eq.~\eqref{eq:qbias_scaling} and the observation that $2k \leq n_0$ for $k \in \Omega_{n_0}$. Furthermore, defining $m' = \lfloor (c_1 - n_0)/ n \rfloor$ as the lower bound for the number of aliases of any $k \in \{ \ell \in \Omega_n \backslash \Omega_{n_0}: c_1 \leq \ell < c_2\}$ we have
\begin{align}
\sum_{k \in \Omega_{n}\backslash \Omega_{n_0}}\frac{\sum_{\ell \in S(k)} \nu_\ell^2}{\left(\sum_{\ell \in S(k) } \nu_\ell \right)^{2}} &\leq \frac{(m+1) (db^2 + 2B_0 ab)^2}{ \left(m' \left[ (d- 2B_0 - k) b^2 + 2B_0 ab \right] +  (m + 1 - m') b^2 \right)^2}
   \\&\leq \frac{n - n_0}{(m+1)}\frac{ (db + 2B_0 a)^2}{ \left(\frac{m'}{m + 1} \left[ (d- 2B_0 - c_1) b + 2B_0 a \right] +  \left(1 - \frac{m'}{m + 1}\right) b \right)^2} \label{line:varub3}
   \\&= \mathcal{O}\left(\frac{n^2}{d}\right), \label{line:varub4}
\end{align}
where in line~\eqref{line:varub3} we have split the denominator into frequencies $n_0/2 \leq k < c_1$ and $c_1 \leq k < d$, bounding $\nu_k$ for the latter as $b^2$. Line~\eqref{line:varub4} follows from observing that $m' = \Theta(d/n) \Rightarrow m' / m = \Theta(1)$, and thus the entire second term of line~\eqref{line:varub3} is bounded as $\mathcal{O}(1)$. Combining Eqs.~\eqref{line:varub2.5} and \eqref{line:varub4} with Eq.~\eqref{eq:var_v2} for the variance of the Fourier features model gives
\begin{equation}
    \textsc{var} = \mathcal{O}\left(\frac{1}{n} + \frac{n}{d} \right) .\label{eq:qvar_scaling}
\end{equation}

It follows that the minimum-$\norm{\cdot}_F$ interpolating quantum model with state preparation unitary $U$ satisfying Eq.~\eqref{eq:gamma_hat} (up to permutations) and data encoded using the Binary encoding strategy will result in benign overfitting as long as the dimensionality of the model scales as $d = \omega(n)$ (i.e., the number of qubits satisfies $n_q = \omega(\log n)$), in which case Eqs.~\eqref{eq:qbias_scaling} and \eqref{eq:qvar_scaling} characterizing the generalization error of the model both vanish in the large-$n$ limit.

A convenience of this demonstration of benign overfitting in a quantum model was that the state $|\Gamma\rangle$ of Eq.~\eqref{eq:gamma_hat} incorporates knowledge of the target function (namely the band-limited spectrum $\Omega_{n_0}$). This could be considered a limitation of the approach, as it imposes an inductive bias on the resulting interpolating model $f^{opt}$, in contrast to other examples of benign overfitting that are more agnostic to the underlying distribution \cite{bartlett2020benign,tsigler2020}. Future work could reveal choices of $|\Gamma\rangle$ that are more data-independent (no explicit dependence on $n_0$) but give rise to \fweight{s} $\nu_k^{opt}$ with the same desirable properties as Eq.~\eqref{eq:nu_bo}. As described in Appendix~\ref{sec:general_bo}, the desirable properties include a more weight on all $\nu_k$ with $k \in \Omega_{n_0}$ and a long, thin tail of \fweight{s} for all other $k \in \Omega_d\backslash \Omega_{n_0}$.

\end{document}